\documentclass[11pt]{article}
\pdfoutput=1
\usepackage{graphicx}
\usepackage{subfigure}
\usepackage{amssymb,amsmath}
\usepackage{slashed}
\usepackage{cite,color,xcolor,url,cancel,ulem}
\usepackage{bm}
\usepackage{jheppub}
\usepackage{color}
\newcommand{\fbinv}{\text{fb}$^{-1} \,$}
\newcommand{\beq}{\begin{equation}}
\newcommand{\eeq}{\end{equation}}
\newcommand{\bea}{\begin{eqnarray}}
\newcommand{\eea}{\end{eqnarray}}
\def\mueff{\mu_\mathrm{eff}}
\def\tanb{\tan\beta}
\def\abprime{A_b^\prime}
\def\atprime{A_t^\prime}
\def\muprime{\mu^\prime}
\def\costhetab{\cos \theta_{\tilde{b}}}
\def\sQ3{\widetilde{Q}_3}
\def\sU3{\widetilde{U}_3}
\def\sD3{\widetilde{D}_3}
\def\stleft{\tilde{t}_L}
\def\stright{\tilde{t}_R}
\def\sbleft{\tilde{b}_L}
\def\sbright{\tilde{b}_R}
\def\stopi{\tilde{t}_i}
\def\stone{\tilde{t}_1}

\def\sboti{\tilde{b}_i}
\def\sbone{\tilde{b}_1}
\def\sbtwo{\tilde{b}_2}
\def\bino{\widetilde{B}}
\def\wino{\widetilde{W}}

\def\higgsinod{\widetilde{H}^0_d}
\def\higgsinou{\widetilde{H}^0_u}

\def\ntrlone{\chi_1^0}
\def\ntrltwo{\chi_2^0}

\def\ntrlonetwo{\chi_{1,2}^0}

\def\charpm{\chi^\pm}

\def\charonep{\chi_1^+}
\def\charonem{\chi_1^-}
\def\charonepm{\chi_1^\pm}

\def\mone{M_1}
\def\mtwo{M_2}

\def\msQthree{m_{\widetilde{Q}_3}}
\def\msUthree{m_{\widetilde{U}_3}}
\def\msDthree{m_{\widetilde{D}_3}}
\def\mstleft{m_{\tilde{t}_L}}
\def\mstright{m_{\tilde{t}_R}}
\def\msbleft{m_{\tilde{b}_L}}
\def\msbright{m_{\tilde{b}_R}}
\def\mstone{m_{\tilde{t}_1}}
\def\msttwo{m_{\tilde{t}_2}}
\def\msbone{m_{\tilde{b}_1}}
\def\msbtwo{m_{\tilde{b}_2}}
\def\mntrlone{m_{{_{\chi}}_1^0}}

\def\mcharone{m_{{_{\chi}}_1^\pm}}

\def\thetasbot{\theta_{\tilde{b}}}

\def\etmiss{\slashed{E}_T}

\title{Sbottoms as probes to MSSM with nonholomorphic soft interactions}
\author[a]{Utpal Chattopadhyay,}
\affiliation[a]{School of Physical Sciences, Indian Association for the Cultivation of Science,\\  
	2A \& B Raja S.C. Mullick Road, Jadavpur, 
	Kolkata 700 032, India}
\author[b]{AseshKrishna Datta,}
\affiliation[b]{Harish-Chandra Research Institute,
HBNI, Chhatnag Road, Jhusi, Allahabad-211019, India}
\author[a]{Samadrita Mukherjee}
\author[a]{Abhaya Kumar Swain}
\emailAdd{tpuc@iacs.res.in}
\emailAdd{asesh@hri.res.in}
\emailAdd{tpsm9@iacs.res.in}
\emailAdd{tpaks@iacs.res.in}
\abstract{Presence of nonholomorphic soft SUSY breaking terms is known to be a
possibility in the popular setup of the Minimal Supersymmetric Standard Model
(MSSM). It has been shown that such a scenario known as NonHolomorphic
Supersymmetric Standard Model (NHSSM) could remain `natural' ({\it i.e.}, not
fine-tuned) even in the presence of a rather heavy higgsino-like LSP. However,
it turns out that distinguishing such a scenario from the MSSM is unlikely to be
an easy task, in particular at the Large Hadron Collider (LHC). In a first study
of such a scenario at colliders (LHC), we explore a possible way that focuses on
the sbottom phenomenology. This exploits the usual $\tanb$-dependence
(enhancement) of the bottom Yukawa coupling but reinforced/altered in the
presence of non-vanishing nonholomorphic soft trilinear parameter $\abprime$.
For a given set of masses of the sbottom(s) and the light electroweakinos (LSP,
lighter chargino etc.) which are known from experiments, the difference between
the two scenarios could manifest itself via event rate in the
2$b$-jets + $\etmiss$ final state, which could be characteristically different
from its MSSM expectation. Impact on the phenomenology of the stops at
the LHC is also touched upon.
}
\keywords{Beyond Standard Model, Supersymmetry Phenomenology, Collider Physics}
\begin{document}
\begin{flushright}
HRI-P-18-09-001
\end{flushright}
\maketitle
\section{Introduction}
\label{Introduction}
The discovery of Higgs boson at the Large Hadron Collider (LHC) of CERN with
mass 125.09 $\pm$ 0.21 (stat.) $\pm$ 0.11 (syst.) GeV
\cite{Aad:2012tfa, Chatrchyan:2012xdj} established the Standard Model (SM) of
particle physics as one of the most successful scientific pursuits of mankind.    
There are, however, a host of theoretical issues along with some experimental
facts that cannot be addressed while staying within the SM. The gauge hierarchy
problem, baryogenesis, the existence of mass for neutrinos, the absence of a particle
dark matter (DM) candidate are a few of the important issues that motivate us
to explore scenarios of physics Beyond the SM (BSM). Supersymmetry (SUSY)
\cite{Nilles:1983ge, Lykken:1996xt, Wess:1992cp, Drees:2004jm, Baer:2006rs,
Haber:1984rc, Martin:1997ns, Chung:2003fi}
remains to be the most attractive framework for pursuing BSM physics.

The fact that the Higgs boson has been found to have a mass well within the
upper limit ($\sim 135$~GeV) predicted by the Minimal Supersymmetric Standard
Model (MSSM) \cite{Drees:2004jm, Baer:2006rs} is something hard to ignore. On the one
hand, this has pushed the lower bound on the masses of the hypothesized SUSY
excitations to higher values, a fact that is especially true for models
employing simpler mechanisms to break SUSY. Such scenarios, thus,
require a greater degree of fine-tuning and hence are less `natural'
\cite{Ellis:1986yg, Barbieri:1987fn, Barbieri:2000gf, Giudice:2013nak}. On the
other hand, even after the 13 TeV run of the LHC, a definitive signature of SUSY
is yet to be established \cite{atlas_link,cms_link}. The null observations have
for sure put serious constraints on various SUSY models. These
together prompt one to explore generic SUSY models with low fine-tuning 
\cite{deCarlos:1993rbr, Kitano:2005wc,
Cabrera:2008tj, Ghilencea:2012qk, Ghilencea:2013hpa, Ghilencea:2013fka,
Baer:2012up, Baer:2013gva, CahillRowley:2012rv, CahillRowley:2012kx,
Cahill-Rowley:2014boa, Perelstein:2007nx, Boehm:2013gst, Girardello:1981wz,
Chan:1997bi}
and looking for their signals in unexplored and perhaps, not so conventional
channels. Such models thus may still survive the latest experimental constraints, but
could become sensitive to ongoing and future LHC searches. Additionally,
models that may have interesting implications for the precision observables,
flavor physics related signatures or those that could offer a suitable DM
candidate would undoubtedly be worth pursuing.  

In the MSSM, SUSY breaking is realized by soft SUSY breaking interactions where,
apart from the mass terms, all the other terms are holomorphic in nature.
However, there is hardly any restriction to include appropriate nonholomorphic
(NH) SUSY breaking interactions 
\cite{Girardello:1981wz, Martin:1999hc, Chattopadhyay:2017qvh, Haber:2007dj,
Bagger:1993ji, Ellwanger:1983mg, Jack:1999ud}. 
It was shown in several works 
\cite{Martin:1999hc, Haber:2007dj, Jack:nh1, Hetherington:2001bk} 
that in the absence of any gauge singlet fields, certain nonholomorphic SUSY
breaking terms may become characteristically soft so that the MSSM may
additionally include soft terms like
$A_f^{\prime}\phi^2\phi^*$ and $\mu^{\prime} \psi\psi$\footnote{There can be
other possible NH terms also like $\lambda \psi$ \cite{Martin:1999hc}. However,
such a term with chiral fermion and gaugino mass mixing is not allowed in the MSSM 
particle content.}. The resulting scenario is broadly known as the
NonHolomorphic Supersymmetric Standard Model (NHSSM).

Implications of the NH soft terms had been analyzed in scenarios with
universal boundary conditions at a high scale
\cite{Sabanci:2008qp, Solmaz2009, Un:2014afa2, Ross:2016pml, Ross:2017kjc}.
These used renormalization group (RG) evolutions of various input parameters to
the weak scale for studying the
SUSY spectra, impact on the Higgs boson mass, involved fine-tuning, various
phenomenological observables in rare $B$-decays ({\it viz.},
BR$(B \rightarrow X_s + \gamma)$, BR$(B_s \rightarrow \mu^+ \mu^-)$), issues
pertaining to the dark matter, CP-violating effects etc.
Analyses that were inspired by the so-called phenomenological MSSM (pMSSM)-like
setup \cite{Djouadi:1998di} (where the input values of the soft SUSY breaking
parameters are provided at the electroweak scale instead) are presented in
\cite{Chattopadhyay:2016ivr, Beuria:2017gtf}. The first of these two works
\cite{Chattopadhyay:2016ivr} demonstrated the enhancement of muon $g-$2 in
models that involve NH terms. The latter one \cite{Beuria:2017gtf} made a
detailed study on the Charge and Color Breaking (CCB) vacua while including
appropriate NH soft breaking terms.
It is noted  that in a formulation of hidden sector $F$-type SUSY breaking, such
NH terms may be suppressed by the scale of mediation \cite{Martin:1999hc}.
Thus, supergravity \cite{Chamseddine:1982jx, Barbieri:1982eh, Hall:1983iz,
Nath:1983aw, Ohta:1982wn} scenarios have NH soft terms that are suppressed by
the Planck mass. Recently, authors of reference \cite{Chattopadhyay:2017qvh} analyzed
an NH scenario in a minimal Gauge Mediated SUSY Breaking (mGMSB) framework
\cite{Drees:2004jm, Giudice:1998bp} where the
scale of mediation is reasonably small. However, like all the analyses to date with
NH terms, we will remain agnostic about the source/mechanism of such
suppressions and consider the parameters (at the electroweak scale) associated
with NH terms to have similar strengths as that of the holomorphic soft
parameters\footnote{Non-standard SUSY breaking was also studied with $R$-parity
violating NH soft SUSY breaking terms in the MSSM framework in references
\cite{Jack:2004dv, Hambye:2000zs, Chakrabortty:2011zz}.}. 

It is now known \cite{Chattopadhyay:2016ivr} that the NH soft trilinear
parameters ($A_f'$) may significantly alter the left-right (L-R) mixing of
squarks and sleptons while the NH soft higgsino mass term ($\muprime$) may
induce crucial changes in the compositions of the electroweakinos.
In particular, as we shall see in section \ref{sec:model}, the masses
of the higgsino-like neutralinos would go as $\mu+\muprime$ at the tree level.
The fact that the
Higgs potential in the NHSSM receives a contribution from the superpotential
parameter `$\mu$' (as in the MSSM), but not from $\muprime$, gives rise to the
possibility that a scenario with relatively heavy higgsino-like neutralinos
could still have a small enough `$\mu$' \cite{Chattopadhyay:2016ivr} and hence a
low degree of tree-level electroweak fine-tuning
\cite{Chan:1997bi, Perelstein:2007nx, Feng:1999zg, Akula:2011jx,
CahillRowley:2012rv, Baer:2012up, Baer:2012cf, Delgado:2014vha,
Mustafayev:2014lqa, Casas:2014eca, Baer:2015rja}.
Such a heavy higgsino ($\sim$ 1~TeV) could be the
Lightest Supersymmetric Particle (LSP) and a viable DM candidate that could
satisfy the observed relic density from the Planck data \cite{Ade:2015xua}.

As it may be expected, in the presence of soft NH terms, the mass-spectra and
the strength of interaction vertices involving SUSY excitations could undergo
significant changes via the altered mass (and hence the diagonalizing) matrices
of the sfermions and electroweakinos. On the other hand, gauge and/or Yukawa
couplings may have dominant roles in determining the strength of these vertices
depending on the composition of sparticles. For example, as we will see in
section \ref{subsec:coupling-sbot}, the strength of the vertices
$\tilde{b}_{i}$-$b$-$\tilde{\chi}^0_{j}$ and
$\tilde{t}_{i}$-$t$-$\tilde{\chi}^0_{j}$
are mostly powered by $y_b$ and $y_t$, respectively when the involved neutralino
is higgsino-like, while the same are governed by gauge couplings when the
neutralino is a gaugino-dominated one. Furthermore, the coupling
strength of a sfermion to a chargino depends crucially on the chiral admixture
the sfermion possesses. One also finds that, among the NH soft trilinear
parameters $A'_f$ (see section \ref{sec:model}), the lepton (tau)- and
the down (bottom)-type ones ($A'_\tau$ and $A'_b$) might contribute
dominantly to the L-R mixings of scalars \cite{Chattopadhyay:2016ivr} for an
enhanced $\tan\beta$ where $\tan\beta$ is the ratio of \textit{vevs}
$(\frac{v_2}{v_1})$ of the two Higgs doublets. As a result, the NH soft
parameters $A'_{\tau,b}$ could induce more L-R mixing in the stau and the
sbottom sectors, respectively than what is contributed by $A_t^{\prime}$ of
similar strength toward mixing in the stop sector. In this work we are
interested in the phenomenology of the squarks from the third generation, in
particular, the sbottom sector and hence would consider only $\abprime$ which is
non-vanishing.

Notably enough, there is another source in the NHSSM that could alter
the strength of the interaction vertices significantly. The bottom Yukawa
coupling $y_b$ could receive large radiative correction in the presence of
non-vanishing $\abprime$ when $\tanb$ is large. This is over and above the usual
$\tanb$-enhancement that $y_b$ enjoys due to radiative corrections in the MSSM
\cite{Hall:1993gn, Hempfling:1993kv, Carena:1994bv, Pierce:1996zz, Logan:2000cz,
Antusch:2008tf}.
Such an effect
arises in the NHSSM essentially from the coupling of the down-type sfermions
(sbottoms) to the up-type Higgs boson triggered by soft trilinear NH interaction
terms ({\it viz.}, $\abprime$). In contrast, consequences of $A_t'$ in the stop
mass-squared matrix is $\tanb$-suppressed and is practically subdominant in the
background of a large holomorphic trilinear coupling $A_t$ that is required to
obtain the SM-like Higgs boson mass in the right range. Also, the impact of
$A_t'$ on $y_t$ via radiative corrections is small. Given that the NHSSM setup
cannot alter the SM-like Higgs boson mass as obtained in the MSSM any
significantly, the requirement of a relatively large $\tanb$ would still be
intact in the NHSSM. This then implies that the NHSSM-specific effects would,
in general, be more pronounced in the phenomenology of the sbottoms when
compared to that of the stops. There is, however, one subtle exception to this
when (an altered) $y_b$ could take part in the decay of the stop squarks.

In this backdrop, already intensified searches at the LHC for the squarks
from the third generation find added relevance and, from the viewpoint of the
NHSSM, sbottoms take the front seat. Given that the masses and the compositions
of the sbottoms predominantly control its (pair) production cross sections and
decays, a detailed analysis of these by looking into their mutual relationships in
the NHSSM framework would be an important first step in deciphering it at
an actual experiment. We thus undertake an exploratory study of the
pair-production of sbottoms at the ongoing 13 TeV run of the LHC. We prefer a
scenario with relatively small values of higgsino mass parameter
`$\mu$' (appearing in the superpotential) to ensure a `natural' setup. If the
soft NH parameter $\muprime$ is not too large, this could ensure the lighter
neutralinos and the lighter chargino to be higgsino-like thus bringing into the
picture a much altered $y_b$ (characteristic of the NHSSM) in their
interactions with the sbottoms. As a result, rates (cross section times the
effective branching fraction or `yields') in various
final states would start to differ from their MSSM expectations. In the present
study, we adopt the final state with 2$b$-jets + missing transverse energy
($\etmiss$) as the reference one.

The present work is organized as follows. In section \ref{sec:model} we discuss 
the salient aspects of the NHSSM scenario with reference to the nonholomorphic
soft terms, their key impacts on the sfermion and the electroweakino sectors
and on the running of the bottom quark masses (Yukawa couplings). Section
\ref{sec:pheno-sbot} is devoted to a comparative discussion on the nature of the
sbottoms, their decays and the resulting yields in our chosen final state as
found in the NHSSM and the MSSM. A brief discussion on the limited but important
impact of the NHSSM scenario on the search for stops at the LHC is also
included.
In section \ref{sec:conclusions} we conclude.
%
\section{Nonholomorphic soft terms and their phenomenological implications}
\label{sec:model}
%
As discussed in reference \cite{Chattopadhyay:2016ivr}, one may consider specific
nonholomorphic interactions which, in the absence of a gauge singlet field, as
in the case of MSSM, may be regarded as soft SUSY breaking terms. The
nonholomorphic cubic scalar interactions and bilinear fermionic mass terms also
potentially fall into the class of soft SUSY breaking ones. Considering $X$ and
$\Phi$ to  be chiral superfields, the NH soft $D$-term contributions go like
$\frac{1}{M^3} [XX^*\phi^2\phi^*]_D$ and
$\frac{1}{M^3} [XX^*D^\alpha\phi D_\alpha\phi]_D$ and 
lead to nonholomorphic terms ({\it e.g.}, $\phi^2\phi^*$ and $\psi \psi$) in the
Lagrangian \cite{Martin:1999hc}. The coefficients of both $\phi^2\phi^*$ and
$\psi \psi$ terms are proportional to $\frac{|F|^2}{M^3}$, where $|F|$ is the
\textit{vev} of the auxiliary field components of $X$. In a scenario with hidden
sector SUSY breaking, nonholomorphic trilinear terms and higgsino mass term
go as $\sim \frac{m_W^2}{M}$ where $M$ is the scale of mediation of SUSY
breaking and can be as large as the Planck scale ($M_P$). Thus, the NH soft
terms are highly suppressed in supergravity type scenarios. In contrast, such
terms can become phenomenologically relevant \cite{Chattopadhyay:2017qvh} in a
scenario like the mGMSB with a
characteristically low scale of mediation of SUSY breaking. As mentioned in the
Introduction, considering the fact that the NH interactions could cause a
significant change in phenomenology, we consider such terms in a
model-independent way so that they can have strengths similar to that of the
holomorphic soft terms.

The parts of the Lagrangian containing the NH terms are given by
\begin{equation}
\begin{split}
\label{nh_lagrangian}
 -\mathcal{L'}_{soft}^{\phi^{2}\phi^{*}} & = \tilde{q}\cdot h_{d}^* {\bf A_{u}'}\tilde{u}^* + 
 \tilde{q}\cdot h_{u}^* {\bf A_{d}'}\tilde{d}^* +\tilde{\ell}\cdot h_{u}^*{\bf
A_{\ell}'}\tilde{e}^* + \mathrm{h.c.},\\
 -\mathcal{L'}_{soft}^{\psi\psi} &={\bf \mu^{\prime}} \tilde{h}_u\cdot
\tilde{h}_d \; ,
\end{split}
\end{equation}
where $A'_{u,d,\ell}$ are the NH soft trilinear parameters corresponding to
the up, down and lepton sectors, respectively and $\muprime$ is the NH soft higgsino
mass parameter. We note that the association of the up and down types of
Higgs fields are reversed when compared to the holomorphic trilinear terms so as
to have the correct hypercharge assignments.

Presence of the NH soft trilinear terms containing $A'_f$ modifies the
off-diagonal terms of the tree-level sfermion mass-squared matrices of the MSSM.
Thus, in the NHSSM, the generic tree-level mass-squared matrix for the sfermions
is given by
\begin{eqnarray}
\label{eqn:sfermion_mass}
M_{\tilde{f}}^2=&\left(\begin{matrix}
m_{\tilde{f_L}}^2+(\frac{1}{2}-\frac{2}{3}\sin^2\theta_{W})m_Z^2\cos2\beta +m_f^2&\hspace{2mm}
	       -m_f(A_f-(\mu+A_f') R_\beta) \\
		-m_f (A_f-(\mu+A_f') R_\beta)  & \hspace{-6mm}
		m_{\tilde{f_R}}^2+\frac{2}{3}\sin^2\theta_{W} m_Z^2 \cos2\beta +m_f^2 
	\end{matrix} \right) \, , \hspace{6mm}
\end{eqnarray}
where $m_{\tilde{f}_{L,R}}$ are the soft SUSY breaking masses for the
left- and the right-chiral sfermions, $m_f$ is the mass of the corresponding
fermion, $A_f$ are the corresponding soft trilinear parameters while `$\mu$' is
the holomorphic higgsino mass parameter appearing in the superpotential.
$R_\beta=\cot\beta \, (\tanb)$ for the up-type squark (down-type squark,
slepton) mass-squared matrices.

As can be seen, when compared to the MSSM case,
`$\mu$' in the off-diagonal term is now replaced by $\mu+A_f'$. At this point, we
must take due note of the fact that, unlike in the MSSM, a contribution from the
trilinear (NH) soft term $A_f'$ appears in a product with $R_\beta$ and hence
could get enhanced by $\tanb$ for the tau and the bottom sectors.
It may also be noted that, in the viable range of
$A_t'$, it is possible to obtain some NH contribution to the radiative
corrections to the mass of the SM-like Higgs boson in spite of a $\tan\beta$
suppression \cite{Chattopadhyay:2016ivr}.

The electroweakino sector is the other entity that gets affected at the tree
level in the presence of NH terms. When compared to the MSSM case, the entries
in the higgsino block of the neutralino and the chargino mass matrices undergo
the modification $\mu \to \mu+\mu'$ \cite{Chattopadhyay:2016ivr} as given by
\begin{eqnarray}
	\label{eqn:neutralino_mass}
	M_{\chi^0}=&\left(\begin{matrix}
		M_1 & 0 & -m_Z\cos\beta \sin\theta_W & m_Z\sin\beta \sin\theta_W \\
		0 & M_2 & m_Z\cos\beta \cos\theta_W & -m_Z\sin\beta \cos\theta_W \\
		-m_Z\cos\beta \sin\theta_W & m_Z\cos\beta \cos\theta_W & 0 & -(\mu+\mu')\\
		m_Z\sin\beta \sin\theta_W & -m_Z\sin\beta \cos\theta_W & -(\mu+\mu') & 0
	\end{matrix} \right) \, , \hspace{6mm}
\label{eqn:ewino}
\end{eqnarray}
\begin{eqnarray}
	\label{eqn:chargino_mass}
	M_{\charpm}=&\left(\begin{matrix}
		M_2 & \sqrt{2} m_W \sin\beta \\
		\sqrt{2} m_W \cos\beta & \mu+\muprime
	\end{matrix} \right) \, , \hspace{6mm}
\label{eqn:ewino}
\end{eqnarray}
where $\mone$ and $\mtwo$ are the $U(1)$ and the $SU(2)$ soft gaugino masses,
respectively.

Here, it is important to note that, at the tree level, the NH soft higgsino
mass parameter $\mu'$ does not contribute to the scalar potential and hence does
not directly affect the mass of the SM-like Higgs boson. As a result, the
issue of electroweak fine-tuning arising from the so-called `little
hierarchy problem' \cite{Barbieri:2000gf, Giudice:2013nak} gets somewhat decoupled from the 
effective higgsino mass given by $\mu+\muprime$ \cite{Chattopadhyay:2016ivr}.
Consequently, one could have a heavy higgsino LSP ($\sim 1$ TeV) (which could be
a viable DM candidate) without compromising on `naturalness' when $\muprime$ is
large enough. 
%
\begin{figure}[t]
     \begin{center}
        \subfigure[]{%
            \label{fig:loopdiagsA}
            \includegraphics[width=0.40\textwidth]{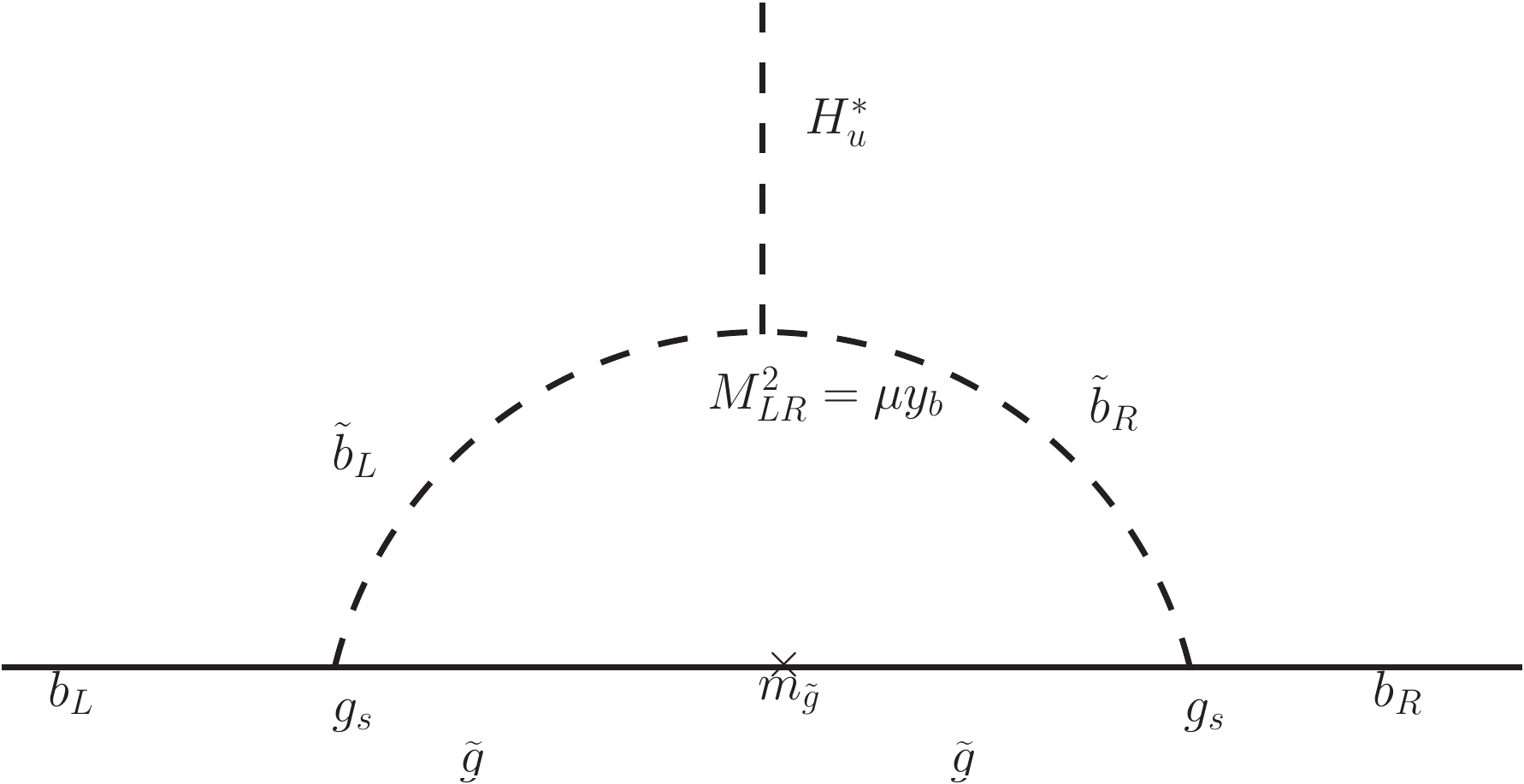}
        }%
\hskip 30pt
        \subfigure[]{%
           \label{fig:loopdiagsB}
           \includegraphics[width=0.40\textwidth]{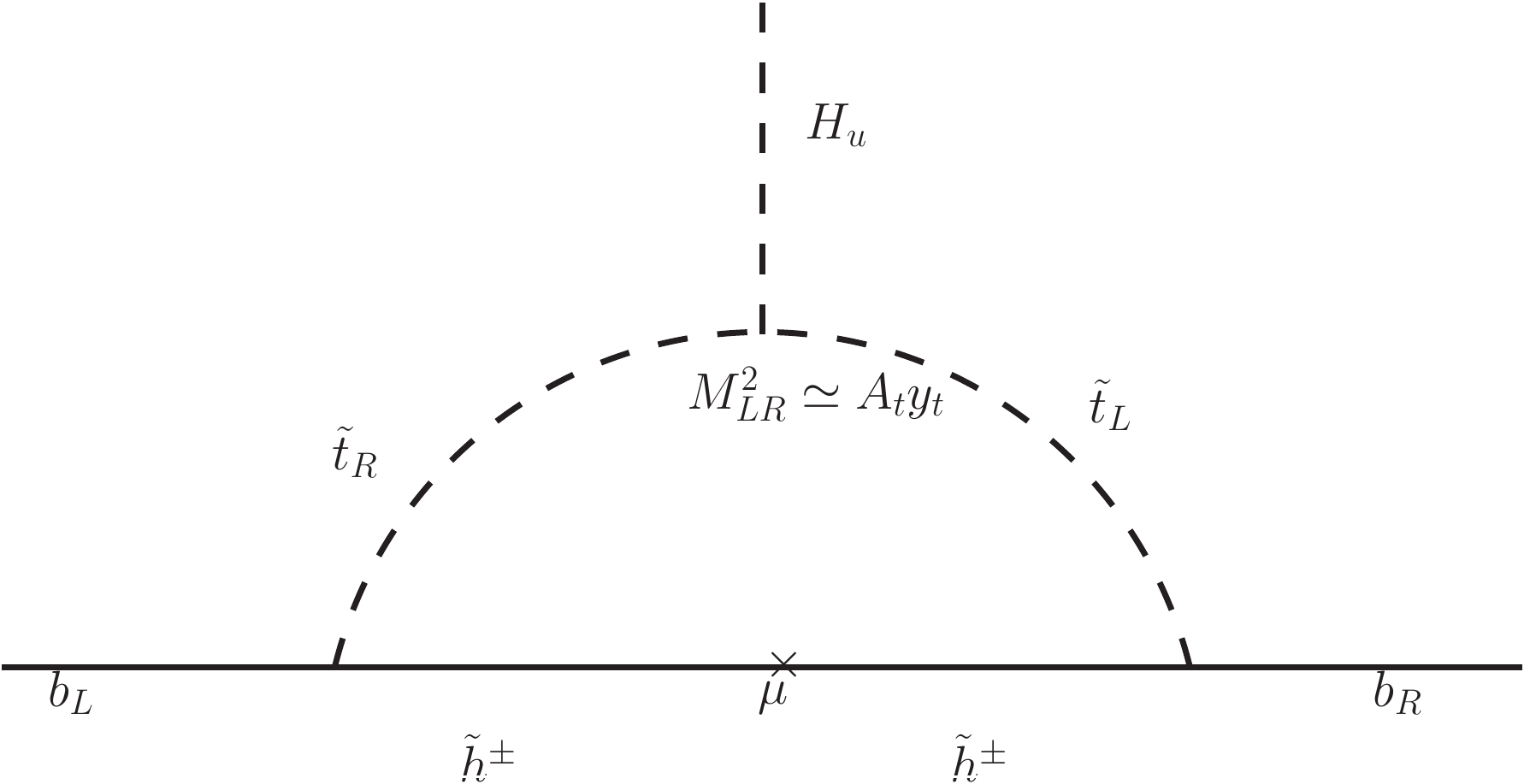}
        }%
        \\
\vspace{0.75cm}
        \subfigure[]{%
           \label{fig:loopdiagsC}
           \includegraphics[width=0.40\textwidth]{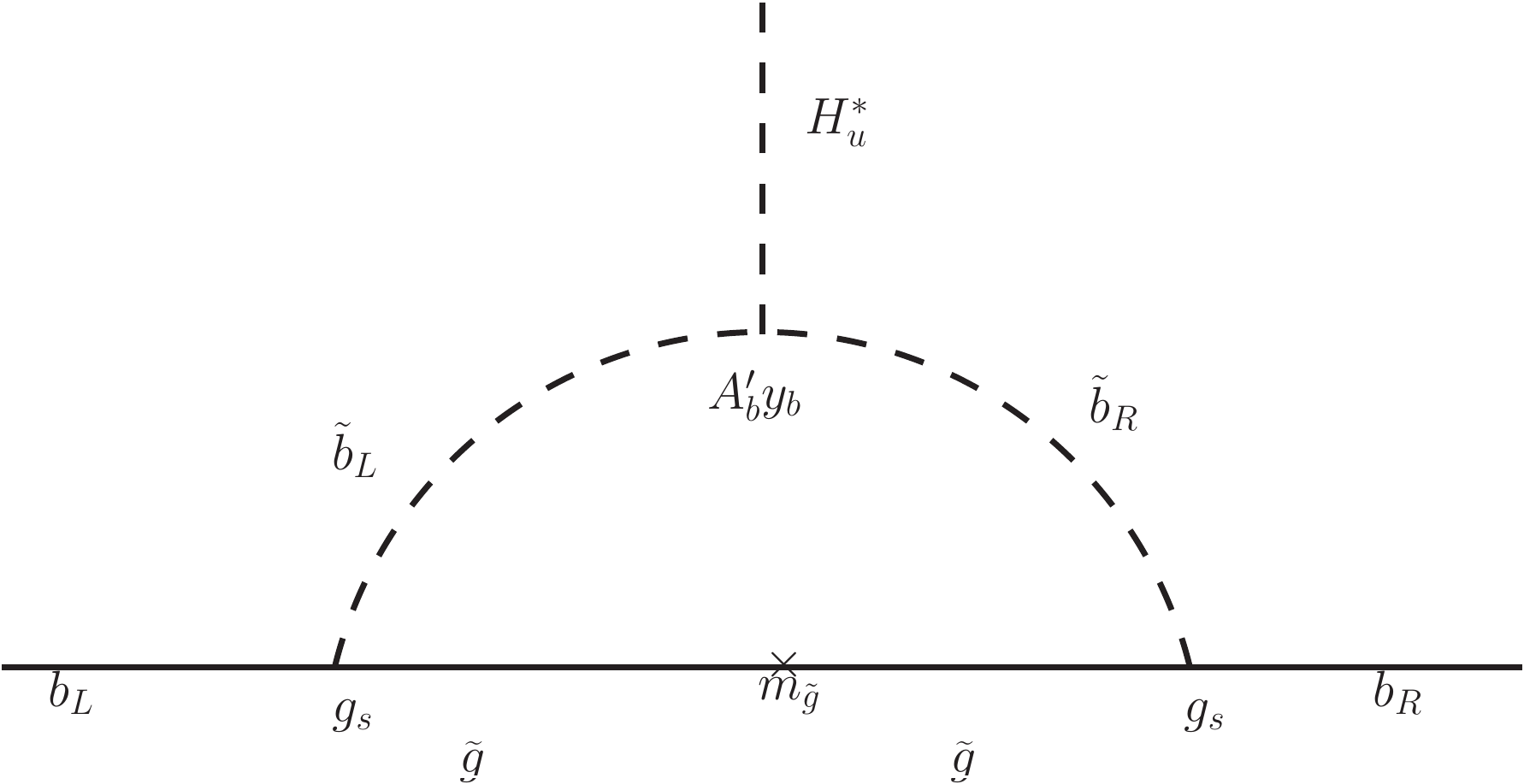}
        }%
\hskip 30pt
        \subfigure[]{%
           \label{fig:loopdiagsD}
           \includegraphics[width=0.40\textwidth]{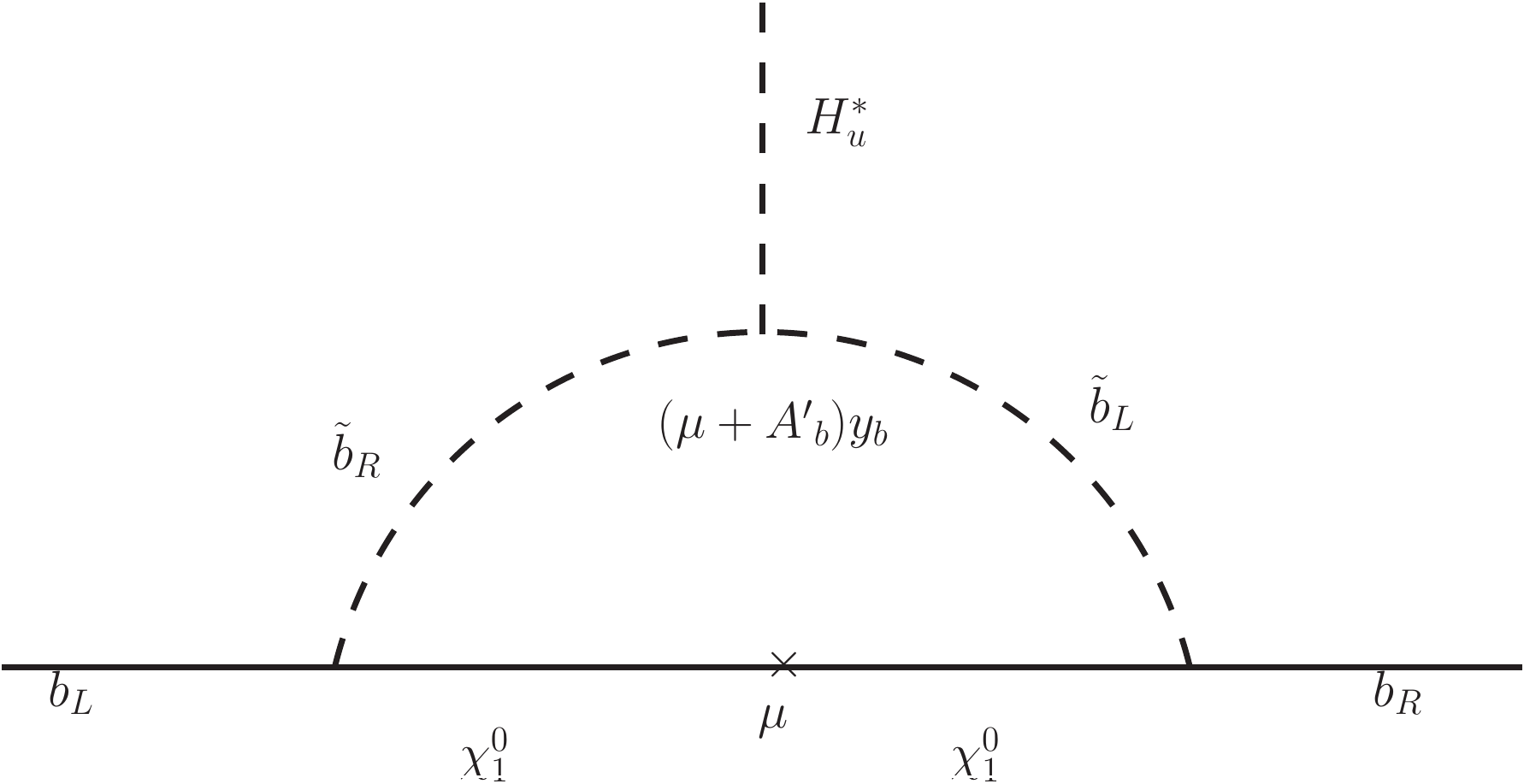}
        }%
        \caption{Principal one-loop diagrams for $\Delta m_b$
          in the MSSM ((a) and (b)) and the NHSSM ((c) and (d)).}
\label{fig:loopdiags}
\end{center}
\end{figure}
%

As pointed out in the Introduction, another significant phenomenological
implication of the NH soft terms is a possible modification in the running
of the fermion mass (Yukawa coupling), in particular, that of the
bottom quark. This is primarily driven by $\abprime$ and further assisted by an
enhanced $\tanb$.  At the one-loop level, the principal MSSM contributions to
the bottom quark mass $m_b$ arise from the $\tilde{g}-\tilde{b}$ loop and the
$\charpm-\tilde{t}$ loop diagrams as shown in figures \ref{fig:loopdiagsA} and
\ref{fig:loopdiagsB} \cite{Hall:1993gn, Hempfling:1993kv, Carena:1994bv,
Pierce:1996zz}.      
The trilinear soft interaction corresponding to the vertex
$H_u$-$\tilde{t}_R$-$\tilde{t}_L$ contributes to the chargino loop via a factor
$A_t y_t v_u$ where $v_u=\langle H_u \rangle$.
The two $q$-$\tilde{q}$-$\tilde{H}$ vertices in this one loop diagram that come from
the superpotential interactions provide with a combined factor of $y_t y_b$.  
For the $\tilde{\chi}^0-\tilde{b}$ loop contribution to $m_b$ in the
MSSM, one can have a trilinear 
soft interaction vertex factor 
$A_b y_b v_d$ in addition to the factor of $y_b^2$ contributed by the
superpotential terms at the other two vertices. 
Thus, the $\chi_0-\tilde b$ contribution in the MSSM 
is much smaller compared to the $\charpm-\tilde{t}$
loop contribution. The $\tilde g-\tilde b$ contribution in the MSSM involves a
factor containing the strong coupling constant $\alpha_s$. This is in addition to the 
factor that goes as $\mu y_b$ at the $\tilde{b}_L$-$H_u^*$-$\tilde{b}_R$ vertex
and arises from the $F$-term contribution to the scalar potential.

On the other hand, in the NHSSM where $H_u$ is associated with the down type squarks
(and sleptons)
in the trilinear soft interactions, one can have a $\tilde g-\tilde b$
contribution with $A_b' y_b$ showing up at the $\tilde{b}_L$-$H_u^*$-$\tilde{b}_R$
vertex (figure \ref{fig:loopdiagsC}).
Besides, there is a $\chi_0-\tilde b$ loop contribution  
(figure \ref{fig:loopdiagsD}) with the same factor of $A_b' y_b$ at the
trilinear scalar vertex and a factor of $y_b^2$
coming from the other two vertices with their origins in the
superpotential. In addition,
we will also take into account the MSSM contribution 
originating from the trilinear vertex involving the squarks and the Higgs boson 
that arises from the
$F$-term of the scalar potential. 

The relevant loop contributions to $m_b$ that we are concerned about in this
work are those which are $\tanb$-enhanced. For the MSSM case, these come from
$\tilde{g}-\tilde{b}$ and $\charpm-\tilde{t}$ loops and are given by
\cite{Hall:1993gn, Hempfling:1993kv, Carena:1994bv, Pierce:1996zz,
Logan:2000cz, Antusch:2008tf}
\begin{eqnarray}
{\Delta m_b^{(\tilde g)}}_{\rm MSSM}&=&\frac{2\alpha_3}{3\pi}m_{\tilde g} \mu y_b 
\frac{v_u}{\sqrt 2} I( \msbone^2, \msbtwo^2, m_{\tilde g}^2), \nonumber \\
{\Delta m_b^{{{\tilde h}^+}}}_{\rm MSSM} &=& \frac{y_t y_b}{16 \pi^2} \mu A_t y_t \frac{v_u}
{\sqrt 2} I( \mstone^2, \msttwo^2, \mu^2) \, , 
\end{eqnarray}
\noindent
where the loop integral $I(a,b,c)$ is given by
\begin{equation}
I(a,b,c)=-\frac{ab~\ln(a/b)+bc~\ln(b/c)+ca~\ln(c/a)}{(a-b)(b-c)(c-a)}.
\end{equation}  
\noindent
For the NHSSM, such contributions come from $\tilde g-\tilde b$ and
$\chi_0-\tilde b$ loops and are given as follows\footnote{A detailed analysis for the effective Higgs vertices involving 
nonholomorphic $A'$ terms with SUSY QCD corrections and electroweak effects can be found in references 
\cite{Crivellin:2010er, Crivellin:2011jt, Crivellin:2012zz}.}:  
\begin{eqnarray}
  {\Delta m_b^{(\tilde g)}}_{\rm NHSSM}&=&\frac{2\alpha_3}{3\pi}m_{\tilde g}
  {A_b^\prime} y_b 
\frac{v_u}{\sqrt 2} I( \msbone^2, \msbtwo^2, m_{\tilde g}^2), \nonumber \\
     { \Delta m_b^{{{\tilde h}^0}}}_{\rm NHSSM} &=& \frac{y_b^2}{16 \pi^2} \mu
     {(\mu+{A_b^\prime}) y_b} \frac{v_u}
{\sqrt 2} I( \msbone^2, \msbtwo^2, \mu^2). 
\end{eqnarray}
Thus, for $\muprime=0$ and for higgsino-dominated $\chi_1^\pm$ and LSP, we have
the following (approximate) expression for $m_b$ where all the loop contributions are 
proportional to $\tanb$:
\begin{eqnarray}
m_b \approx \frac{y_b v_d}{\sqrt 2}
\Big[1 +
\frac{y_t^2}{16 \pi^2}\mu A_t I( \mstone^2, \msttwo^2, \mu^2) \tan\beta +
   \frac{2\alpha_3}{3\pi}m_{\tilde g} (\mu + A_b^\prime)
  I( \msbone^2, \msbtwo^2, m_{\tilde g}^2)\tan\beta
\nonumber \\
 + \frac{y_b^2}{16 \pi^2}\mu (\mu+A_b^\prime) I( \msbone^2, \msbtwo^2, 
\mu^2) \tan\beta \Big]. 
\end{eqnarray}
For a higgsino-dominated lighter chargino and the LSP, and for non-vanishing
$\mu^\prime$, the loop function $I(a, b, c)$ will take ${(\mu+\mu^\prime)}^2$ in its
argument in the NHSSM, instead of $\mu^2$.
%
\section{Sbottom phenomenology in the NHSSM (vis-a-vis the MSSM)}
\label{sec:pheno-sbot}
%
As discussed earlier, the couplings involving the physical sbottom
states and the electroweakinos are of particular importance in the study
of how the NHSSM scenario could behave differently from the MSSM scenario.
We thus first write down the expressions for the tree-level couplings involving
these states. As can be seen, these couplings are functions of the elements
of the matrices that diagonalize the mass (mass-squared) matrices in the
electroweakino (scalar) sectors. As for the electroweakinos, the issue is
straight-forward, {\it i.e.}, these elements are mostly governed by $\mueff=\mu+\muprime$
and the gaugino mass parameters like $\mone$ and $\mtwo$. On the other hand, for
the sbottom (and stop, as well) sector the elements of the diagonalizing matrix
are functions of the nonholomorphic parameter $\abprime \, (\atprime)$ along
with the standard MSSM parameters like $A_b$, `$\mu$' and $\tanb$. 

It needs to be pointed out here that, given the way $\muprime$ enters
the mass matrices in the electroweakino sector at the tree level (see equations
\ref{eqn:neutralino_mass} and \ref{eqn:chargino_mass}), it would be difficult to
decipher an NHSSM effect just by studying the charginos and/or the neutralinos
in isolation. Structurally, the same is true when dealing with the sbottom
sector, given the
way $\abprime$ appears in the mass-squared matrix. When considered exclusively,
both sectors could be viewed as though endowed with an effective value of `$\mu$'
given by $\mu+\muprime$ for the electroweakino sector and $\mu+\abprime$ for the
sbottom sector, respectively. On their own, the mass-matrices would then mimic
their MSSM counterparts. However, any study of sbottoms at the colliders would
inevitably involve their eventual decays to the LSP and possibly, to other lighter
electroweakinos.
It thus appears that definite imprints of the NHSSM
could only be carried by the direct interactions of the sbottoms with the
electroweakinos.
%
\subsection{Interactions with electroweakinos}
\label{subsec:coupling-sbot}
%
With the decay processes $\sboti \to b \ntrlone$ and $\sboti \to t \charonem$ in
reference where $i=1 (2)$ refer to the lighter (heavier) mass eigenstates of the
sbottom, their transition matrix elements involve the couplings
$\sboti$-$b$-$\ntrlone$ and $\sboti$-$t$-$\charonem$, respectively. These
contain a generic factor like $C_L P_L + C_R P_R$,
where $P_{L,R}= {{1 \mp \gamma_5} \over 2}$ are the usual projection operators
and $C_L$ and $C_R$ for the two processes are given by
\cite{Kraml:1999qd, Staub:2013tta, Staub:2015kfa}
\begin{itemize}
\item for $\sboti$-$b$-$\tilde{\chi}^0_1$ coupling:
\begin{equation}
\begin{split}
  C_L &=  -\frac{i}{6} (-3\sqrt{2} g_2 N_{12}^{*} Z^d_{i3} + 6 N_{13} y_b
Z^d_{i6} + \sqrt{2} g_1 N_{11} Z^d_{i3}),  \\
  C_R &= -\frac{i}{3} (3 y_b Z^d_{i3} N_{13} + \sqrt{2} g_1
Z^d_{i6} N_{11}),
\end{split}
\label{eqn:sbbn1}
\end{equation}
\item for $\sboti$-$t$-$\tilde{\chi}^{-}_1$ coupling:
\begin{equation}
\begin{split}
 C_L &= i(y_t Z^d_{i3} V_{12}), \\
 C_R &= i(-g_2 U_{11}^* Z^d_{i3} + U_{12}^* y_b Z^d_{i6}).
\end{split}
\label{eqn:sbtc1}
\end{equation}
\end{itemize}
In the above expressions, $N_{ij}$ are the elements of the symmetric
($4\times 4)$ matrix that diagonalizes the neutralino mass matrix of the same
dimension in the basis ($\bino, \, \wino, \, \higgsinod, \, \higgsinou$).
$U_{ij}$ and $V_{ij}$ are the two unitary matrices that diagonalize the
($2\times 2$) asymmetric chargino mass matrix in the basis of charged wino and
charged higgsino states. $Z^d_{i3}$ and $Z^d_{i6}$ are the elements of the
unitary matrix that diagonalizes the mass-squared matrix for the down-type 
quark that give the left- and right-sbottom admixtures respectively, in the
`$i$'-th sbottom mass-eigenstate. $g_1$ and $g_2$ are the usual $SU(2)$ and
$U(1)$ gauge couplings, respectively, while $y_b$ and $y_t$ are the bottom and
the top Yukawa couplings, respectively. For all these, we follow the convention
and the notations adopted in the {\tt SARAH (v4.10.2)}
\cite{Staub:2013tta, Staub:2015kfa} and {\tt SPheno (v4.0.3)} \cite{Porod:2011nf}
packages which take $0\leq \thetasbot \leq \pi$. Also, if $m_{LL}^2 < m_{RR}^2$,
$\costhetab > {1 \over \sqrt{2}}$ and $\sbone$ would have a dominant $\sbleft$
admixture where $m_{LL}$ ($m_{RR}$) is the diagonal entry in equation
\ref{eqn:sfermion_mass} for the left (right) sectors, respectively. The converse
is true as well.

At this point, it is important to note that the coupling of a sbottom state to a
higgsino-like neutralino is always proportional to $y_b$ while the same to a top
quark and a higgsino-like chargino depends on the chiral admixture it possesses.
Such a coupling for a left-like sbottom is governed by $y_t$ while that for a
right-like sbottom goes as $y_b$. Hence, clearly, when allowed by phase space, a
left-like sbottom dominantly decays to $t\charonem$ thus leading to a small
branching fraction for the $b \ntrlonetwo$ final state when $\ntrlonetwo$ are 
both higgsino-dominated and light. This is no different from the MSSM. However,
in the NHSSM scenario, the presence of a non-vanishing $\abprime$
alters the composition of the sbottom states in a nontrivial way. This could
lead to different observable effects when compared to the MSSM case.
Note that there may be another competing decay mode of an sbottom state in the
form of $\sbone \to \stone W^-$ which is often considered in a detailed study
of the sbottom sector. In the present work, however, we restrict ourselves to a
simpler situation where this decay is kinematically forbidden.

\begin{figure}[t]
\begin{center}
\includegraphics[width=8.5cm]{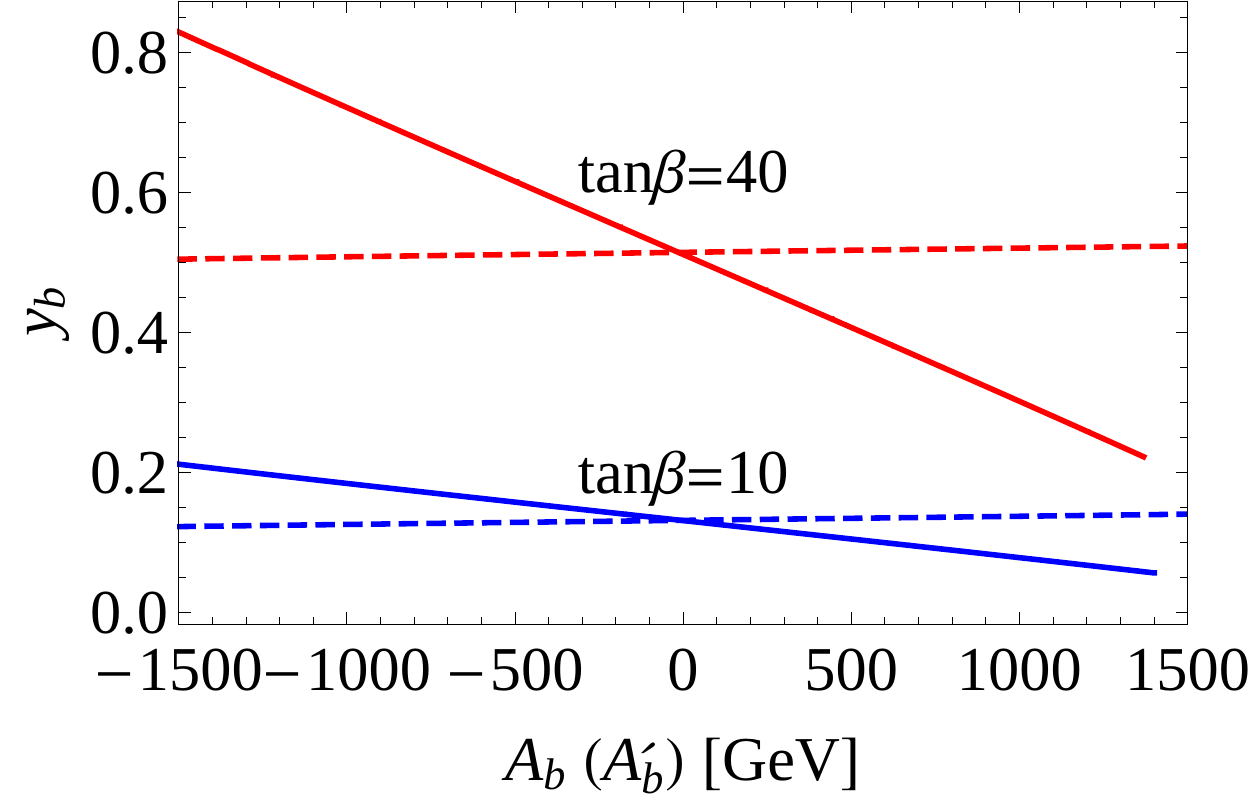}
\caption{Variation of $y_b$ as a function of $\abprime$ (in the NHSSM with
$A_b=0$; bold lines) and $A_b$ (in the MSSM; broken lines) for $\tanb=10$ (in
blue) and for $\tanb=40$ (in red). Some of the fixed input parameters are
$\mu=200$ GeV, $\muprime=0$,  $\mone=500$ GeV and $\mtwo=1.1$ TeV while the rest are presented
in table \ref{tab:inputs}.
}
\label{fig:yb-ab-abprime}
\end{center}
\end{figure}

Interestingly, the couplings mentioned above receive nontrivial contributions
through $y_b$. The effects here are of two types: first, $y_b$ has the usual
dependence on $\tanb$ as in the MSSM case
\cite{Hall:1993gn, Hempfling:1993kv, Carena:1994bv, Pierce:1996zz, Logan:2000cz, Antusch:2008tf}.
More crucially, in the NHSSM scenario, $y_b$ becomes a function of $\abprime$
via radiative corrections to $m_b$. The dependence
is studied via the {\tt SARAH}-generated {\tt SPheno} implementation of the NHSSM scenario
The variation is illustrated in figure \ref{fig:yb-ab-abprime} for two values
of $\tanb$.
The curves reveal usual dependence (direct proportionality) of $y_b$ on
$\tanb$. In addition, significantly enough, the variation of $y_b$ with
$\abprime$ is found to be rather large (up to 4 times for the range of
$\abprime$ we consider). As can be gleaned from figure \ref{fig:yb-ab-abprime},
the larger values of $y_b$ are obtained for larger $\tanb$ and for
$\abprime < 0$. In contrast, the variation
of $y_b$ as a function of $A_b$ in the MSSM is known to be much smaller in
comparison. These are shown in broken lines for the two values of $\tanb$ under
consideration. It is also to be noted that, for a given $\tanb$, NHSSM leads to
larger $y_b$ values when compared to the MSSM case only if $\abprime <0$.
The reverse is the case when $\abprime >0$.
\begin{figure}[h]
\begin{center}
\subfigure{
\includegraphics[width=7.5cm]{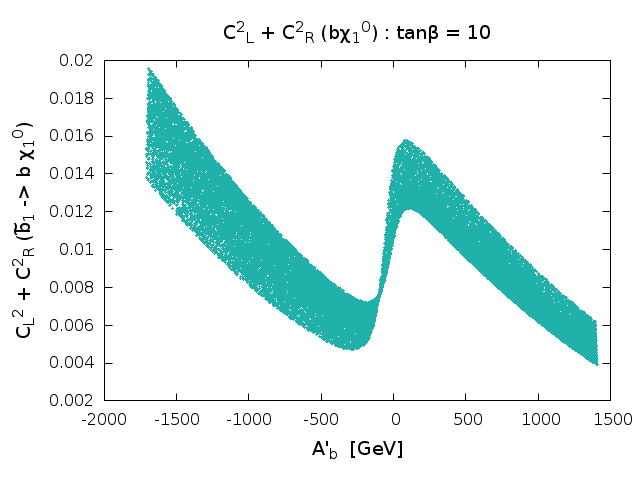}
 }
\subfigure{
\includegraphics[width=7.5cm]{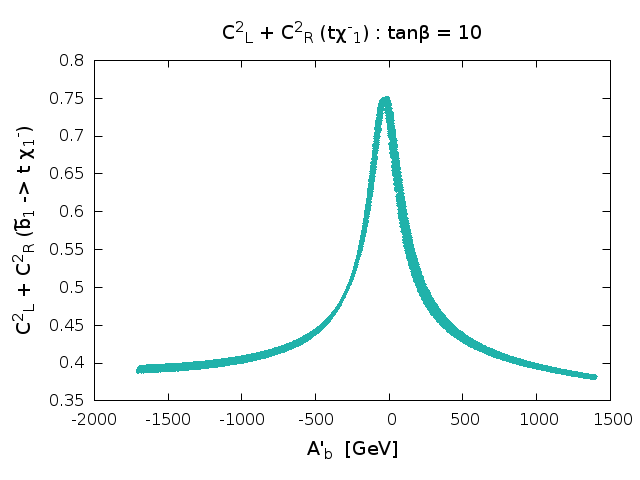}
 }
\subfigure{
\includegraphics[width=7.5cm]{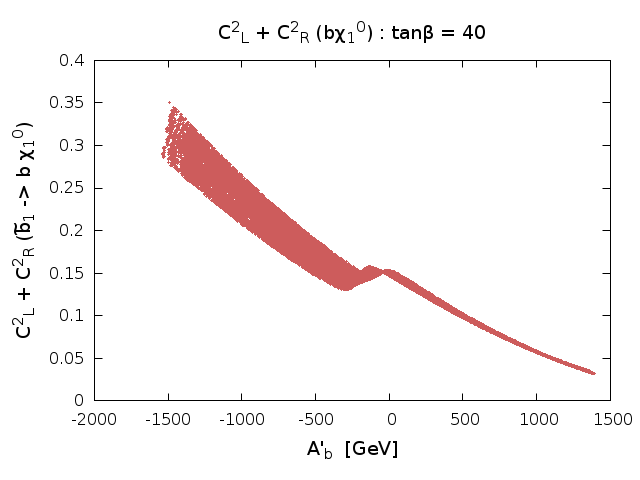}
 }
\subfigure{
\includegraphics[width=7.5cm]{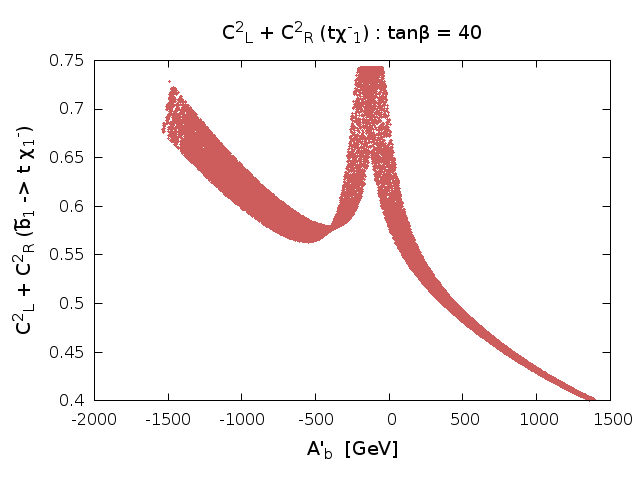}
 }
\caption{Variation of the effective interaction strengths ($C_L^2+C_R^2$) 
involved in the decays $\sbone \to b \ntrlone$ (left) and
$\sbone \to t \charonem$ (right) as functions of $\abprime$ for
$\tanb=10$ (top) and for $\tanb=40$ (bottom) in a scenario with a higgsino-like
neutralino LSP. The fixed input range of `$\mu$' is $100$ GeV
$\leq \mu \leq 350$ GeV. Other fixed parameters used are
as follows: $\mone=500$ GeV, $\mtwo=1.1$ TeV, $\muprime=0$, $A_b=0$ while
$\msbleft$= $\msbright$=1.2 TeV.}
\label{fig:couplings-abprime-higgsino}
\end{center}
\end{figure}
%
%
\begin{figure}[h]
\begin{center}
\subfigure{
\includegraphics[width=7.5cm]{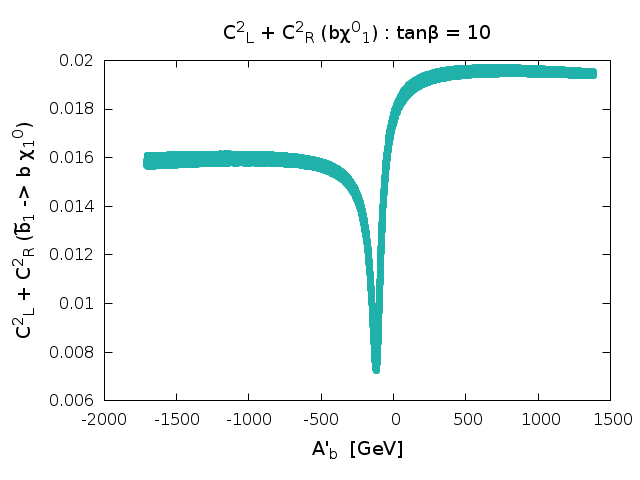}
 }
\subfigure{
\includegraphics[width=7.5cm]{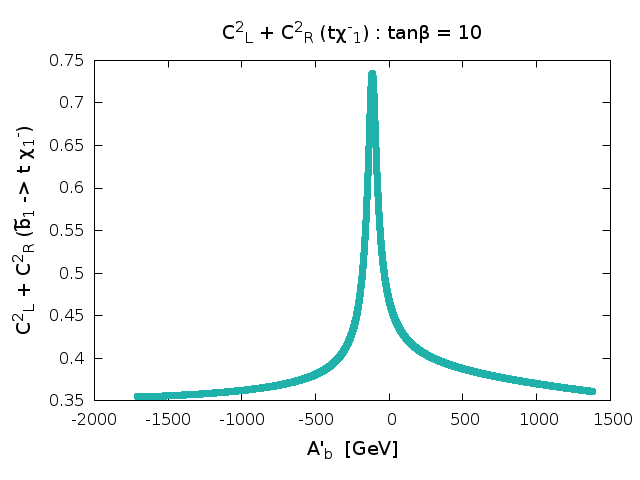}
 }
\subfigure{
\includegraphics[width=7.5cm]{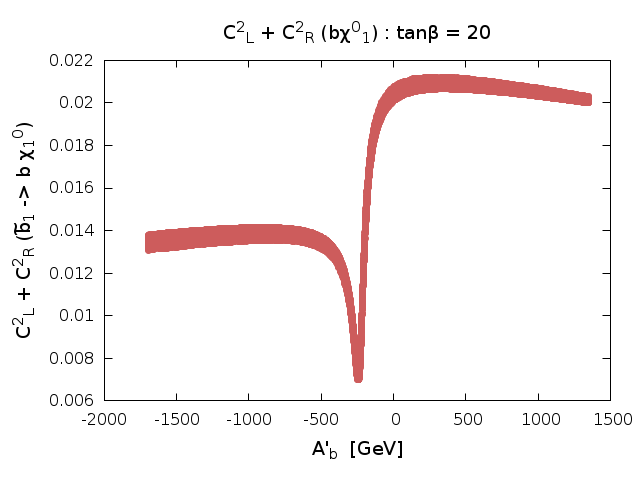}
 }
\subfigure{
\includegraphics[width=7.5cm]{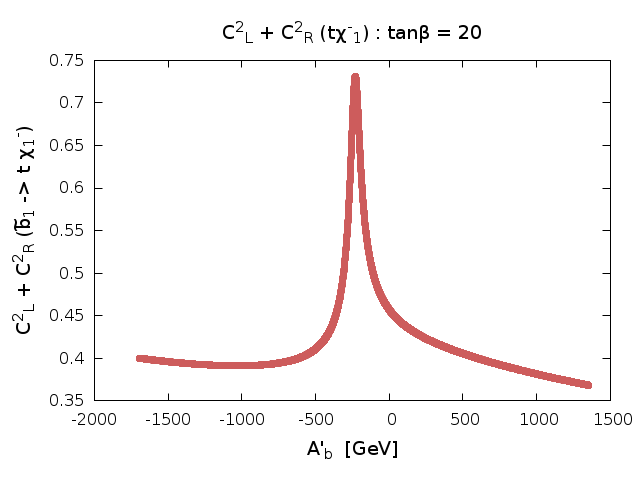}
 }
\caption{Same as in figure \ref{fig:couplings-abprime-higgsino} but for a
scenario having a gaugino (bino)-like LSP with $\mu=900$ GeV while the fixed
input range of $\mone$ used is $100$ GeV $\leq \mone \leq 350$ GeV. Also, the
larger of the two $\tanb$ values used here is 20 (bottom panel) instead of 40
used in figure \ref{fig:couplings-abprime-higgsino} (see text for details).
}
\label{fig:couplings-abprime-gaugino}
\end{center}
\end{figure}
%

We now move on to study the dependence of the effective strengths of the
vertices $\sbone$-$b$-$\ntrlone$ and $\sbone$-$t$-$\charonem$ that appear in the
decays $\sbone \to b \ntrlone$ and $\sbone \to t \charonem$, respectively, on
$\abprime$. This assumes that pair-production of $\sbone$ has the dominant
contribution to the final state we would be interested in. However, in section
\ref{subsubsec:lhc-sbtwo-pair} we would take a careful look as to how
significant the contributions from $\sbtwo$ pair-production could turn out to
be.

At the level of squared matrix elements for these decays, the interaction
strengths are constituted of appropriate $C_L$'s and $C_R$'s of equations
\ref{eqn:sbbn1} and \ref{eqn:sbtc1}. We have checked analytically that squared
matrix-elements for these decays dominantly depend on the factor $C_L^2+C_R^2$.
We thus study the variations of this factor as functions of $\abprime$ and
$\tanb$ for the two decay processes mentioned above and for two representative
scenarios: one with a higgsino-like neutralino LSP
(figure \ref{fig:couplings-abprime-higgsino}) and the other with a gaugino
(bino)-like neutralino LSP
(figure \ref{fig:couplings-abprime-gaugino}). The values of $C_L$ and
$C_R$ for the two decay processes are obtained from {\tt SARAH}.
In both figures, plots on left
(right) stand for the effective strength of the vertex $\sbone$-$b$-$\ntrlone$
($\sbone$-$t$-$\charonem$) and the top panels correspond to an input $\tanb$
value of 10. However, the bottom panel in figure
\ref{fig:couplings-abprime-higgsino}
(figure \ref{fig:couplings-abprime-gaugino}) corresponds to $\tanb=40 \; (20)$.
The compulsion for choosing two different values of `higher' $\tanb$ (40 versus
20) to demonstrate the said variation in the scenarios with higgsino- and
gaugino-like LSP is elaborated later in section \ref{subsec:masses}, in an
appropriate context.

It is clear from figure \ref{fig:couplings-abprime-higgsino} that, for both
modes of decay, the involved interaction strengths vary significantly with
$\abprime$. The bands arise since `$\mu$' is varied (over the range
$100$ GeV $\leq \mu \leq 350$ GeV) which, all through, keeps the LSP dominantly
higgsino-like) for our choices of $\mone$ and $\mtwo$. For the
$\sbone$-$b$-$\ntrlone$ vertex, the effective interaction strength could vary by
a factor as large as 5 (10) for $\tanb=10 \; (40)$. For the
$\sbone$-$t$-$\charonem$ vertex, the corresponding variation is found to be by a
factor of 2, approximately, irrespective of $\tanb$. Also, the generic strength
of the effective $\sbone$-$b$-$\ntrlone$ interaction is much smaller than that
for $\sbone$-$t$-$\charonem$. This is mainly because the latter is governed by
$y_t$ when $\sbone \sim \sbleft$, which happens to be the case here.

The profiles in figure \ref{fig:couplings-abprime-higgsino} can also be
understood in the following terms. For the $\sbone$-$b$-$\ntrlone$ interaction
(plots on the left), the LSP being dominantly a higgsino, the coupling factor
$C_L^2+C_R^2$ is mostly determined by $y_b$. Hence its variations with
$\abprime$ exhibit a nature similar to the corresponding variations of $y_b$.
Dips appear at $\abprime$ values with commensurate magnitudes of `$\mu$' for
which $\mu + \abprime \simeq 0$. This flags a situation with the least
chiral-mixing between the sbottom states thus rendering the $\sbone$ state to be
almost purely left-handed. Thereon, an increase in $\abprime$, as it
changes its sign to positive, flips the sign of $\costhetab= Z^d_{1,3}$ but the
latter still (nearly) retaining the magnitude it had for $\abprime <0$. This
results in shooting up of the interaction strength as the sign on $\abprime$
flips from negative to positive. Further increase in $\abprime$ leads to a
monotonic drop in the interaction strength which merely follows the trend
$y_b$ exhibits as a function of $\abprime$.
We will see later (see figure \ref{fig:couplings-abprime-gaugino}) that this feature gets reflected in the variation
of the branching ratio BR[$\sbone \to b \ntrlone$] as a function of $\abprime$.

For the $\sbone$-$t$-$\charonem$ interaction (plots on the right) issues are a
little more involved. As noted earlier in this section, in the central region,
with $\sbone \sim \sbleft$, the interaction strength for $\sbone$-$t$-$\charonem$
preferentially depends on $y_t$. This effect is reflected in the peaks in the
center of these plots. Further, the bigger the magnitude of $\abprime$ is, the
larger is the admixture of $\sbright$ in $\sbone$ which, in turn, makes the
interaction increasingly more dependent on $y_b$. However, for $\tanb$ not so
large (=10), as is the case in the top right plot, even for large negative
$\abprime$, $y_b$ cannot compete with $y_t$ that shapes the peak around
$\abprime=0$. In contrast, in the bottom right plot, for which $\tanb$ is much
larger (=40), $y_b$ is  expectedly much enhanced for large negative $\abprime$
(see figure \ref{fig:yb-ab-abprime}). This drives the interaction strength up to
the level obtained for the $y_t$-driven regime of the central region with small
$\abprime$.
This plot also corroborates with figure \ref{fig:yb-ab-abprime} that $y_b$
decreases steadily with increasing positive value of $\abprime$. 
The bands appear due to a variation of `$\mu$' over the range indicated.
while the central peak arises from $y_t$-dominance.   
Overall, it may be noted that the effective strength is small for the
$\sbone$-$b$-$\ntrlone$ interaction when compared to the one for the
$\sbone$-$t$-$\charonem$ interaction. In fact, for $\tanb=10$, the former always
remains much smaller.

Descriptions of the plots in figure \ref{fig:couplings-abprime-gaugino} follow
a similar line of argument but for the fact that the LSP is now bino-dominated.
Thus, the $\sbone$-$b$-$\ntrlone$ coupling is driven by the hypercharge of the
sbottom. Hence the interaction strength would be higher for $\sbone$ with a
larger admixture of $\sbright$. This is clearly reflected in the plots on the
left where dips appear at the central region with smaller $|\abprime|$ for which
the $\sbone$ is $\sbleft$-dominated. With an increasing magnitude of $\abprime$,
$\sbright$ admixture grows quickly to a maximum 
(see the $\cos\theta$ variation plot of figures \ref{fig:mass-br-higgsino}
and \ref{fig:mass-br-gaugino}.)
and the effect of this is
reflected in the wings of these curves that take off fast and then spread out
flat for larger values of $|\abprime|$.
On the other hand, plots on the right representing the $\sbone$-$t$-$\charonem$
case (in particular, the top right one with $\tanb=10$) are (is) rather similar
to the corresponding ones (one) in figure \ref{fig:couplings-abprime-higgsino}
since the lighter chargino still remains to be dominantly higgsino-like, though
heavier, as $\mcharone \approx \mu = 900$ GeV and  $\mone << \mu << \mtwo$,
with $\mone$ and $\mtwo$ set at 500 GeV and 1.1 TeV, respectively. Some
differences arise for the bottom right plot for larger magnitudes of $\abprime$.
This is since, as described earlier, in this region the $y_b$ effect dominates
and, given that we are using a smaller value for $\tanb$ (20, versus 40, in
figure \ref{fig:couplings-abprime-higgsino}), $y_b$ is smaller in the present
case, for a given $\abprime$. Note further that the parts in the central region
(showing the peaks) where $y_t$ governs are expectedly not affected much by this
difference in $\tanb$. Again, as for the higgsino-dominated LSP scenario of
figure \ref{fig:couplings-abprime-higgsino}, here also the strength of the
$\sbone$-$b$-$\ntrlone$ interaction remains to be much smaller than that of the
$\sbone$-$t$-$\charonem$ interaction. Also, note that in contrast to figure
\ref{fig:couplings-abprime-higgsino}, even for the larger $\tanb$ (=20) value
considered in this figure, the $\sbone$-$b$-$\ntrlone$ coupling does not get
enhanced since it dominantly involves gauge coupling $g_1$ and not $y_b$. 
%
\subsection{Masses,  mixings and decays of the lighter sbottom}
\label{subsec:masses}
%
The branching fractions of $\sbone$ to various competing modes are governed not
only by the couplings involved but also by the phase space available for such
decays. The latter, in turn, depends on the mass-splittings between $\sbone$ and
the sparticle it decays to, of which the important ones are the electroweakinos
(a neutralino or a chargino) and, possibly, the lighter stop state ($\stone$).
As mentioned earlier, in the present study, decays of $\sbone$ to
electroweakinos would play more direct roles though. Hence, for simplicity, we
consider $\msbone < \mstone + m_W$ such that the decay $\sbone \to \stone W^-$
is kinematically forbidden. 
\begin{figure}[t]
\begin{center}
\subfigure{
\includegraphics[width=7.5cm]{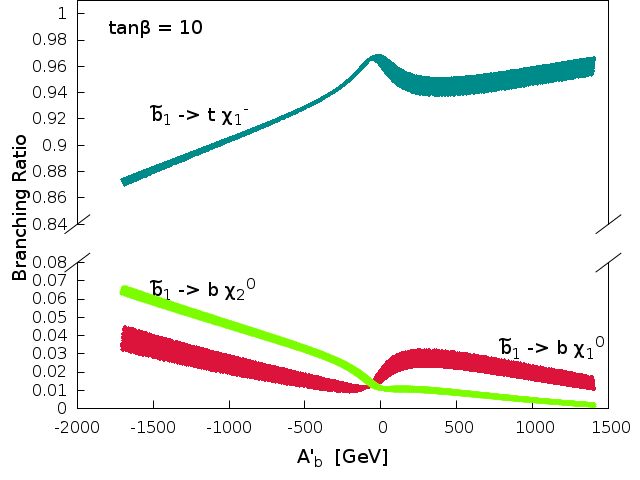}
}
\subfigure{
\includegraphics[width=7.5cm]{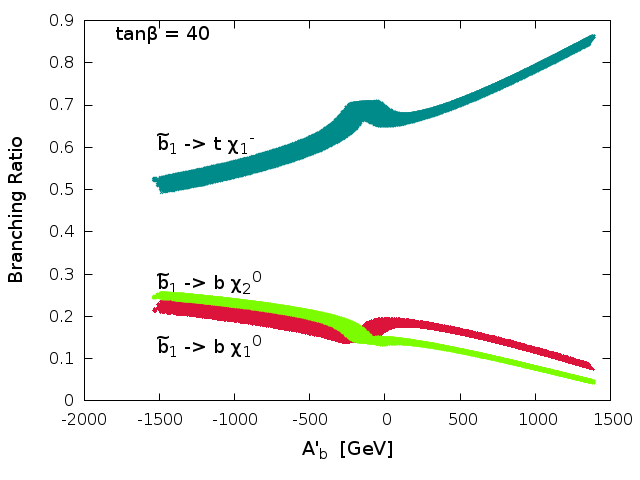}
}
\caption{Variations of various decay branching fractions of $\sbone$ as a
function of $\abprime$ when the LSP is higgsino-like for $\tanb=10$ (left) and
$\tanb=40$ (right). Other fixed/varying input parameters are as in figure
\ref{fig:couplings-abprime-higgsino}. A broken ordinate is chosen to present the
branching ratio in the plot on the left.
}
\label{fig:abp-br-higgsino}
\end{center}
\end{figure}
\begin{figure}[t]
\begin{center}
\subfigure{
\includegraphics[width=7.5cm]{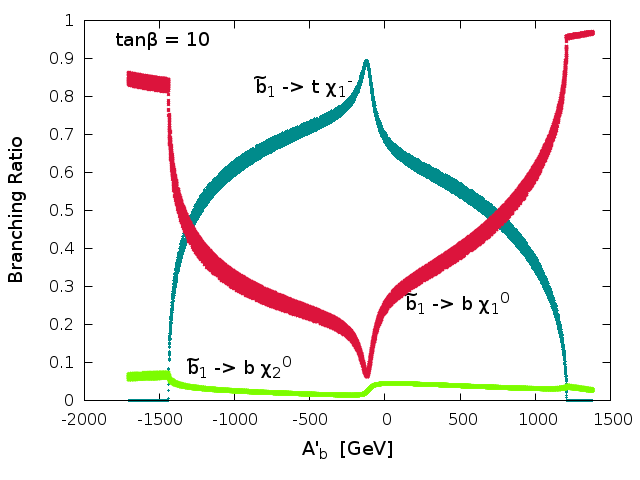}
}
\subfigure{
\includegraphics[width=7.5cm]{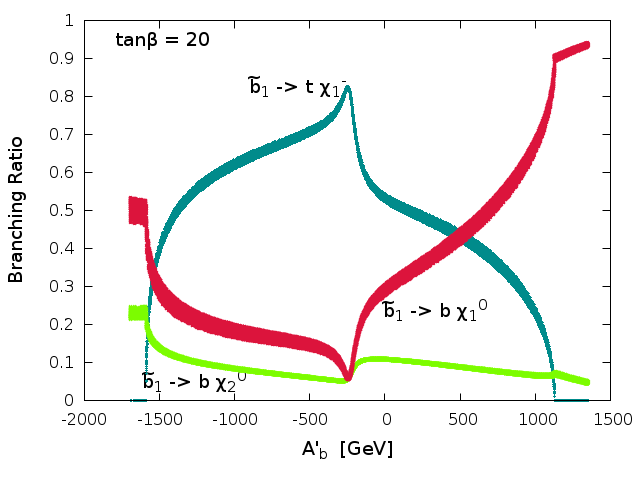}
}
\caption{Same as in figure \ref{fig:abp-br-higgsino} but for a bino-like LSP.
Other fixed/varying parameters are as in figure  
\ref{fig:couplings-abprime-gaugino}.
}
\label{fig:abp-br-gaugino}
\end{center}
\end{figure}

We discuss two representative scenarios in this
context. In the first scenario, $\mueff$ remains small and is much smaller than
$M_{1,2}$. This renders $\ntrlone, \, \ntrltwo$ and $\charonepm$ higgsino-like.
In the second one, $M_1 << \mueff << M_2$ leading to a bino-like neutralino LSP
while $\charonepm$ is still higgsino-like but could be much heavier than the
LSP. In figures \ref{fig:abp-br-higgsino} and
\ref{fig:abp-br-gaugino} we present the variations of the branching fractions
of $\sbone$ as functions of $\abprime$ for these two scenarios, respectively.
In both these figures, plots on the left (right) correspond to $\tanb=10 \,
(40)$. We find that the patterns of variation are very different for the two
scenarios. On a closer look, we find that these derive from those
for the corresponding interaction strengths as depicted in figures
\ref{fig:couplings-abprime-higgsino} and \ref{fig:couplings-abprime-gaugino}.
This can also be understood in the following way. The interaction strength for
$\sbone$-$t$-$\charonem$ is generally larger than that for
$\sbone$-$b$-$\ntrlone$, the difference getting smaller with increasing $y_b$
({\it i.e.}, at large negative $\abprime$ and large $\tanb$).
Hence the total decay width appearing in the
denominator of the formula for branching fractions is dominated by the width
of $\sbone \to t \charonem$ and competes with $\sbone \to b \ntrlonetwo$
only in the region where $y_b$ is enhanced. 
In the regions where the interaction strengths for $\sbone \to t \charonem$ have
flatter profiles ({\it e.g.},
for $\tanb=10$ and larger $|\abprime|$), the variation of BR[$\sbone \to b \ntrlone$]
would follow the same for its decay width which, in turn, goes as the
corresponding interaction strength. For smaller $|\abprime|$ where the change
of the width for $\sbone \to t \charonem$ is somewhat abrupt, the variation of
BR[$\sbone \to b \ntrlone$] still broadly follows the profile of the involved
coupling strength. This is more or less true for both the cases with higgsino-
and gaugino-dominated LSP. The region where $y_b$ becomes large ({\it e.g.}, for
$\tanb=40$ and larger $|\abprime|$) is only relevant when the LSP (lighter
chargino) is higgsino-dominated for which a competition among the said decay
modes set in and the profiles for the branching ratios shape up accordingly. 

\begin{figure}[t]
\begin{center}
\includegraphics[width=7.5cm]{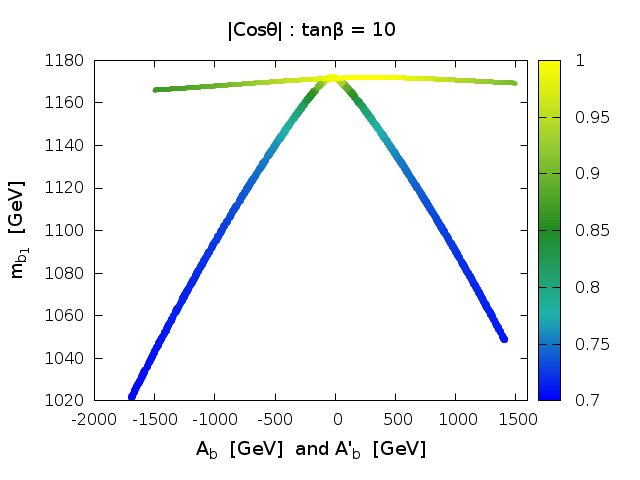}
\includegraphics[width=7.5cm]{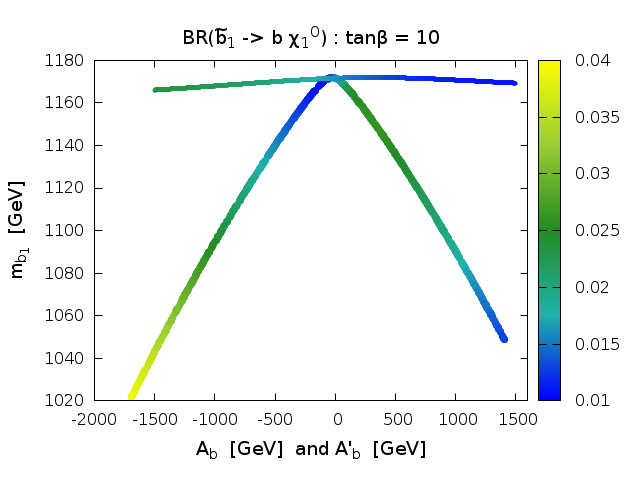}
\vskip 20pt
\includegraphics[width=7.5cm]{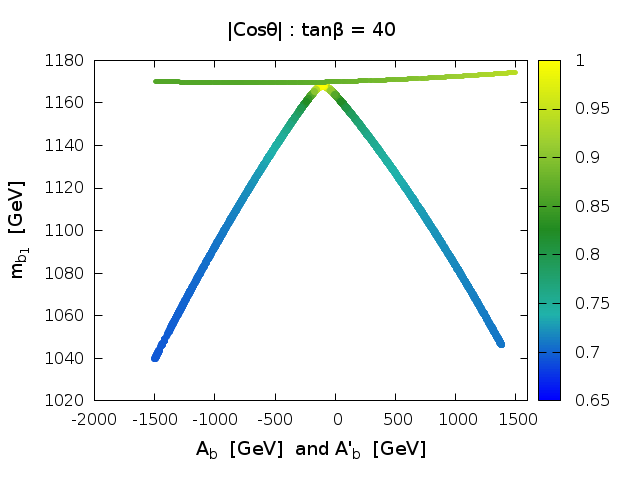}
\includegraphics[width=7.5cm]{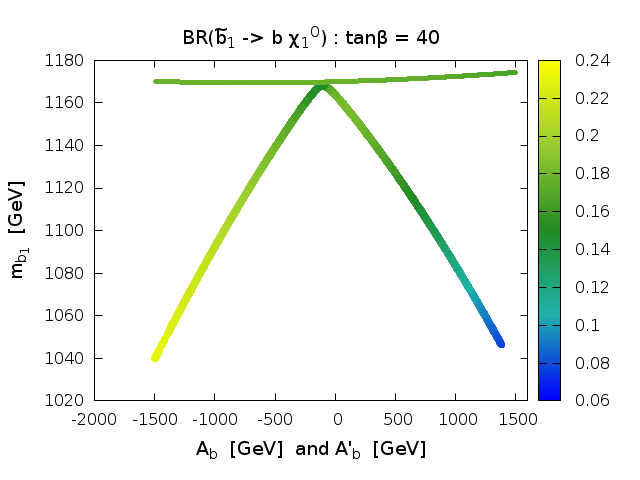}
\end{center}
\caption{Variations of $\costhetab$ (left) and BR[$\sbone \to b \ntrlone$]
(right) in the $\abprime$-$\msbone$ plane for $\tanb=10 (40)$ in the top
(bottom) panel for a scenario with both the LSP neutralino and the lighter
chargino being higgsino-like. The fixed parameters used are as follows:
$\mu=200$ GeV, $\mone=500$ GeV, $\mtwo=1.1$ TeV, $A_b=0$, $\muprime=0$,
$\msQthree$=$\mstleft$= $\msbleft$=$\msbright \, (\msDthree)$=1.2 TeV and
$\mstright \, (\msUthree) =1.5$ TeV. Flatter lines represent the corresponding
variations as functions of $A_b$ in the MSSM.
}
\label{fig:mass-br-higgsino}
\end{figure}

\begin{figure}[t]
\begin{center}
\includegraphics[width=7.5cm]{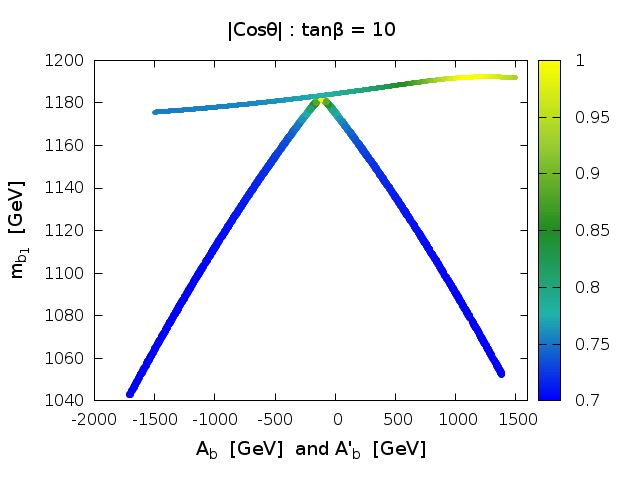}
\includegraphics[width=7.5cm]{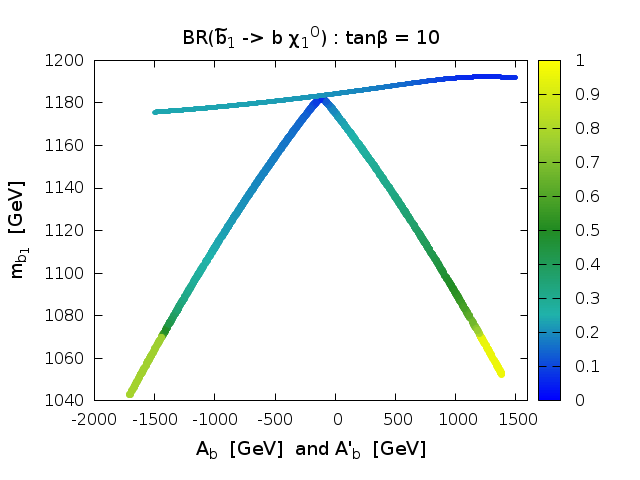}
\includegraphics[width=7.5cm]{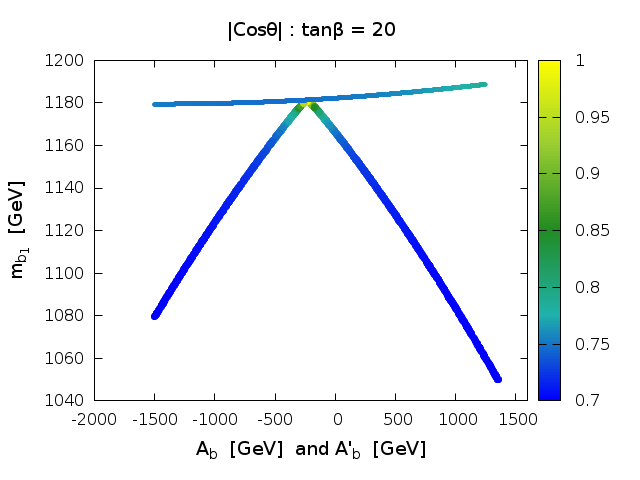}
\includegraphics[width=7.5cm]{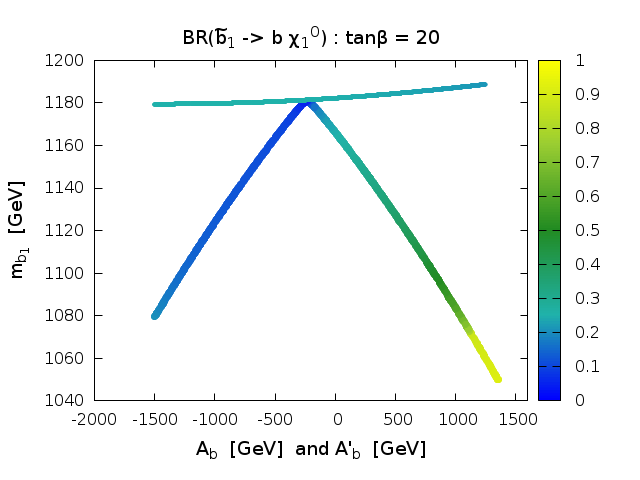}
\end{center}
\caption{Same as in figure \ref{fig:mass-br-higgsino} but for a scenario with
gaugino (bino)-like neutralino LSP which is ensured by setting $\mone=500$ GeV,
$\mtwo=1.1$ TeV and $\mu=900$ GeV while $\tanb=20$ for the plots in the
bottom panel. The specific choice of `$\mu$' also ensures that the decay mode
$\sbone \to t \charonem$ could be open or closed depending upon varying
$\msbone$ as a function of $\abprime$.
\label{fig:mass-br-gaugino}
}
\end{figure}

We further need the information as to how $\msbone$ varies as a function of
$\abprime$ ($A_b$, in the MSSM). For simplicity, this can be studied for a fixed
set of values of the sbottom soft masses ($\msbleft$ and $\msbright$)\footnote{
Throughout this work, we ensure that the parameter points are consistent with
the current bounds on the Higgs sector as implemented in the packages {\tt
HiggsBounds (v4.3.1)} \cite{Bechtle:2013wla} and {\tt HiggsSignals (v1.4.0)}
\cite{Bechtle:2013xfa}.}. 
Intimately connected with this variation is the chiral content of $\sbone$ which,
in turn, dictates the strength of its various interactions.
Hence presenting simultaneous variations of these factors would perhaps be the
best means to understand their interplay. In figures \ref{fig:mass-br-higgsino}
and \ref{fig:mass-br-gaugino} we illustrate these variations for the cases with
a higgsino-like and a gaugino (bino)-like neutralino LSP, respectively.

In both figures \ref{fig:mass-br-higgsino} and \ref{fig:mass-br-gaugino}, the
top panel presents the case for $\tanb=10$. However, for the bottom panel
$\tanb$ is set to 40 (20) for figure \ref{fig:mass-br-higgsino}
(figure \ref{fig:mass-br-gaugino}). The reason behind this is that the larger
value of $\mu \, (=900 \, \mathrm{GeV})$ used to obtain a gaugino-like neutralino
LSP in the latter case hardly finds a situation with $\tanb \gtrsim 20$ and
still yielding an acceptable (non-tachyonic) spectrum of particles over the range
of $\abprime$ that ensures a substantial variation (drop) in $\msbone$.
Note that the plots in the top panel directly correspond to the same in figure
\ref{fig:couplings-abprime-higgsino} depicting the involved coupling strengths.

All plots are shown in the $\abprime (A_b)$-$\msbone$ plane. Thus, the common
backdrop they present is the variation of $\msbone$ as a function of $\abprime$
($A_b$) in the NHSSM (MSSM). The fixed value of the soft masses are taken to be
$\msbleft$= $\msbright$=1.2 TeV. This is so chosen that, for the entire range
of variation of other input parameters, $\msbone$ always remains around 1 TeV
thus more than satisfying the most conservative bound from the LHC
\cite{Aaboud:2017wqg} on the same. The values/ranges we employed for all the
relevant SUSY input parameters are presented in table \ref{tab:inputs}. 
For figure \ref{fig:mass-br-higgsino}
(figure \ref{fig:mass-br-gaugino}) we choose $\mu=200$ (900) GeV while $\mone$ and
$\mtwo$ are fixed at 500 GeV and 1.1 TeV, respectively for both the figures.

For both figures \ref{fig:mass-br-higgsino} and
\ref{fig:mass-br-gaugino}, we have made a simplifying choice of
$\muprime=A_b=0$ for the NHSSM case\footnote{Since $\muprime$ and
$\abprime$ exclusively affect only the electroweakino and the sbottom sectors,
respectively (at least, at the tree-level), $\abprime=0$ could appear to be a
viable alternate choice. However, given that $y_b$ has a non-trivial dependence
on $\abprime$, here we choose to vary the latter. In section
\ref{subsubsec:lhc-sbone-pair} (see figure \ref{fig:muprime-abprime}), we would
simultaneously vary $\muprime$ and $\abprime$ to study some observable effects.
}.
For the MSSM case, $A_b$ is varied over the same range as for $\abprime$ in the
NHSSM case.
Varying colors (from the adjacent palettes) indicate changing sbottom mixing
angle (in terms of $\cos\thetasbot$) for plots on the left of these figures and
varying branching ratio (fraction) BR[$\sbone \to b \ntrlone$] for
those on the right. Clearly, $\costhetab$ ranges between ${1 \over \sqrt{2}}
\approx 0.7$ (maximal mixing) and 1 ($\sbone \equiv \sbleft$ limit) signifying
$\sbone$ to be $\sbleft$-dominated.
Flatter lines at the top of these plots illustrate the
corresponding variations in the MSSM for varying $A_b$ and are introduced to
highlight the extent to which issues basic to the phenomenology (masses and
mixing angles) could get altered in the NHSSM scenario.

Figures \ref{fig:mass-br-higgsino} and \ref{fig:mass-br-gaugino} reveal the
following important information. These tell us that over the range of variation
of $\abprime$ shown, $\msbone$ could vary by $\lesssim 160$ GeV. It is a
significant variation in view of the fact that the corresponding number in the
MSSM (as a function of $A_b$, varied over the same range as $\abprime$) reaches
at most 20 GeV. This could be understood in the following way. It may appear
that a comparable range of variation in $\msbone$, in the MSSM, could be found
just by allowing `$\mu$' to vary over a larger range thereby compensating for
the missing $\abprime$. However, this is not correct. In fact, the major effect
in this respect, in the NHSSM, does not come directly from $\abprime$, per se,
in the off-diagonal element of the mass-squared matrix. Rather, a significant
variation of $y_b$ with $\abprime$, as illustrated in figure
\ref{fig:yb-ab-abprime}, induces such a big change in $\msbone$.
In addition, plots on the left clearly show that, in the NHSSM, with increasing
magnitude of $\abprime$, one quickly achieves a close-to-maximal mixing in the
sbottom sector while the same is difficult to find in the MSSM, for a varying
$A_b$.

Plots on the right of these figures depict the variation of
BR[$\sbone \to b \ntrlone$].
One finds that in a scenario with a higgsino-like LSP (figure
\ref{fig:mass-br-higgsino}), BR[$\sbone \to b \ntrlone$] is typically small and
barely reaches $\sim 25\%$ for large $\tanb$ values while, in the scenario with
a gaugino (bino)-like LSP (figure \ref{fig:mass-br-gaugino}), the same could
attain a value of 100\%. This is grossly a kinematic effect where the larger
value of `$\mu$' leads to a heavier $\charonepm$ thereby suppressing the competing
BR[$\sbone \to t \charonem$]. Also, note that larger values of
BR[$\sbone \to b \ntrlone$] are obtained for $\abprime <0 \; (>0)$ for
scenarios with a higgsino-like (gaugino-like) LSP. However, the reasons behind
such enhancements in the two cases have a subtle difference. For the former,
BR[$\sbone \to b \ntrlone$] gets directly reinforced due to an enhanced $y_b$ as
$\abprime <0$. For the latter, the enhanced BR[$\sbone \to b \ntrlone$] is due
to closure of the decay $\sbone \to t \charonem$ as $\msbone$ drops to a
critical level with increasing value of $\abprime >0$.
It may be noted at this point that a $\sbright$-dominated $\sbone$ would lead
to an enhanced branching fraction to bottom quark and LSP for both the higgsino-
and bino-dominated LSP. Thus, by sticking to  $\sbone \simeq \sbleft$, we made
a conservative choice as far as the final-state yields are concerned.
%
\begin{table}[t]
\centering
\begin{tabular}{|c|c|c|}
\hline\hline 
Parameters  &  MSSM (GeV) &  NHSSM (GeV)  \\ [0.5ex]
\hline
$M_{2,3}$
& \multicolumn{2}{|c|}{1100, 2800} \\
$m_{\tilde{Q}_{1,2}}/ m_{\tilde{D}_{1,2}}/ m_{\tilde{U}_{1,2}}$
& \multicolumn{2}{|c|}{2000} \\
$m_{\tilde{Q}_{3}}, m_{\tilde{D}_{3}}, m_{\tilde{U}_{3}}$
& \multicolumn{2}{|c|}{1200,1200,1500} \\
$m_{\tilde{L}_{1,2}}/m_{\tilde{E}_{1,2}}$
& \multicolumn{2}{|c|}{1800} \\
$m_{\tilde{L}_{3}}/m_{\tilde{E}_{3}}$
& \multicolumn{2}{|c|}{2000} \\
$m_A$
& \multicolumn{2}{|c|}{2000} \\
$\tan\beta$
& \multicolumn{2}{|c|}{10} \\
& \multicolumn{2}{|c|}{} \\
\hline
$A_{t},A_{\tau}$
& \multicolumn{2}{|c|}{2000,0} \\
\cline{2-3}
$A_{b}$ & [-1500 : 1500] & 0 \\
$\abprime$  & -- & [-1500 : 1500]  \\
\cline{2-3}
$A_{t}',A_{\mu}',A_{\tau}'$
& -- & {0,0,0} \\
$\mu'$  & -- & [-1500 : 1500] \\
 & & \\
\hline\hline 
Observables &  MSSM (GeV) &  NHSSM (GeV)  \\ [0.5ex]
\hline
$m_{\tilde g}$
& \multicolumn{2}{|c|}{2709} \\
\cline{2-3}
$m_{\tilde t_1},m_{\tilde t_2}$
& 1120.2, 2087.2 & 1103.5, 2087.2 \\
$m_h$ & 125.1 & 125.3 \\
\cline{2-3}
$m_H$, $m_{H^{\pm}}$
& \multicolumn{2}{|c|}{1978, 1979} \\
& \multicolumn{2}{|c|}{} \\
\hline
\hline
${\rm BR}(B \rightarrow X_s + \gamma)$
& $3.30 \times 10^{-4}$ & $3.50 \times 10^{-4}$ \\
${\rm BR}(B_s \rightarrow \mu^+ \mu^-)$
& $3.20 \times 10^{-9}$ & $3.02 \times 10^{-9}$\\
\hline
\end{tabular}
\caption{Fixed input parameters used in this work (unless otherwise specified)
for the MSSM and the NHSSM scenarios. Values for the masses and the trilinear
coupling parameters are shown in GeV. In both scenarios we set $\mu = 200$ GeV
(900 GeV) to ensure a higgsino (gaugino)-like neutralino LSP. Chosen values of
$\tanb$, $\mone$ and $\mtwo$ are indicated in the text in individual contexts.
{\tt SARAH (v4.10.2)}-generated {\tt SPheno (v4.0.3)} has been used to generate
the particle spectra and for the estimation of the branching ratios and other
observables. Parameter points are checked to be consistent with the current
bounds as implemented in the packages {\tt HiggsBounds (v4.3.1)} and
{\tt HiggsSignals (v1.4.0)}.
}
\label{tab:inputs}
\end{table}
%
\subsection{Pair-productions of sbottoms at the LHC}
\label{subsec:lhc-pheno}
%
In this section we study the effect of the NHSSM-specific parameters like
$\muprime$ and $\abprime$ on the (parton-level) signal strengths
($\sim \sigma \times \mathrm{BR}^2$) of sbottoms produced in pairs at the LHC and
each decaying to a bottom quark and an LSP. This leads to a final state with
2$b$-jets + missing transverse energy ($\etmiss$) and is being
intensively searched for at the LHC experiments leading to the strongest lower
bound on the sbottom mass. Hence, as discussed in the Introduction,
this might prove to be an appropriate process to find imprints of a scenario
like the NHSSM. These could be in the form of yield in the above-mentioned final
state that is different from the MSSM expectation, for a given set of
masses of sbottoms, the lighter chargino and/or the lighter neutralinos.
We, thus, study the relative yields in these two SUSY scenarios. This we do in
stages. In section \ref{subsubsec:lhc-sbone-pair} we stick to $\sbone$
pair-production and consider cases with vanishing and non-zero $\muprime$.
$\sbtwo$ pair-production is additionally considered in section
\ref{subsubsec:lhc-sbtwo-pair} (for the simpler case with vanishing $\muprime$)
to demonstrate the situation when it cannot be ignored. 

\subsubsection{The case with the lighter sbottom}
\label{subsubsec:lhc-sbone-pair}
It may be reiterated that, at the lowest order in the perturbation theory, the
NHSSM parameter $\muprime$ affects only the electroweakino sector while the
other NHSSM parameter in context, $\abprime$, does so only for the sbottom
sector of the said scenario. In contrast, `$\mu$' affects both the sectors and
does so in both MSSM and NHSSM. Furthermore, as we have already discussed in
section \ref{sec:model}, $\tanb$ could also play an important role. Clearly,
letting all the relevant parameters vary simultaneously and still be able to
extract some concrete information is a difficult proposition. We, thus, need an
appropriate strategy to obtain information which, if not free from, has less of
an ambiguity and thus could be used to decipher an imprint of the NHSSM scenario
in the experimental data. Note that we would, all through (unless otherwise
specified), assume $\msbleft$= $\msbright$=1.2 TeV which approximately sets the
magnitude of $\msbone$ to be around 1 TeV (thus, evading the current bounds from
the LHC) even in presence of maximal mixing between the two chiral states.
All through we consider a vanishing $A_b$ (unless otherwise mentioned) given
that it is found to play only a subdominant role in the sbottom sector. 
\begin{figure}[t]
\begin{center}
\subfigure{
 \includegraphics[width=7.5cm]{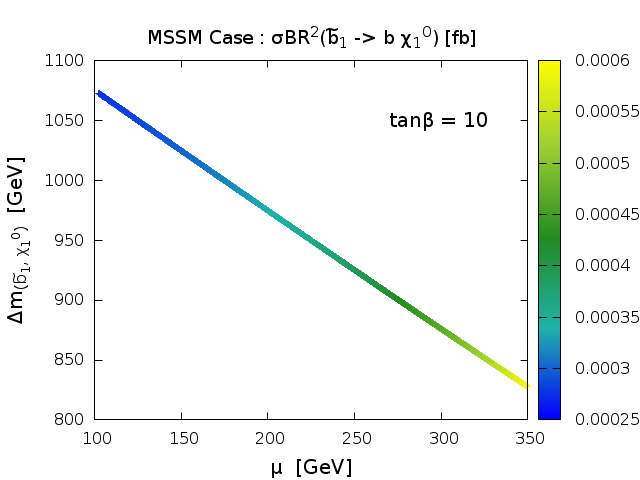}
  }
  \subfigure{
	\includegraphics[width=7.5cm]{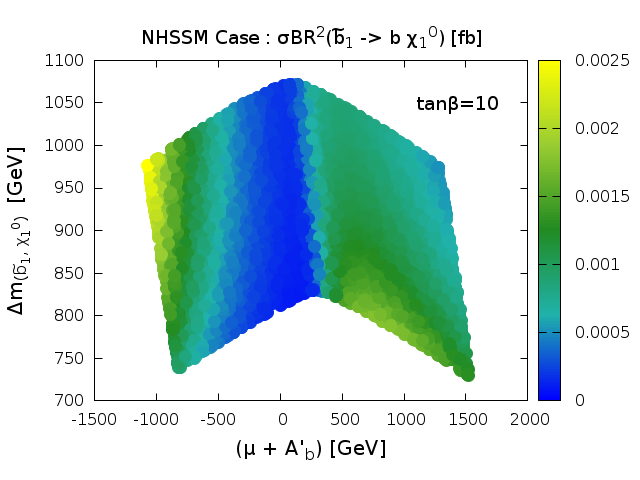}
	}
	\subfigure{
	\includegraphics[width=7.5cm]{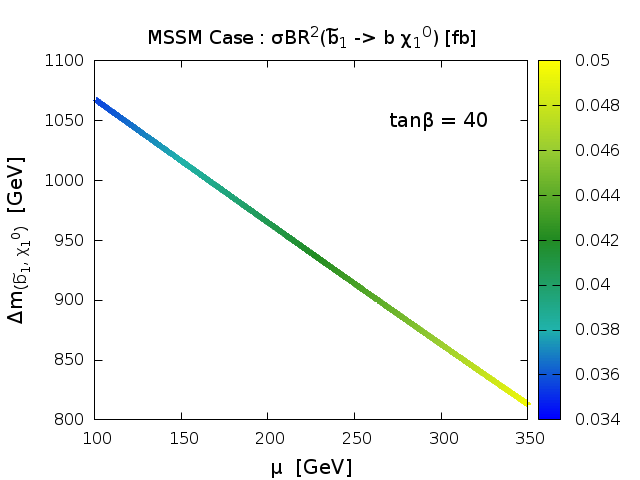}
	}
	\subfigure{
	\includegraphics[width=7.5cm]{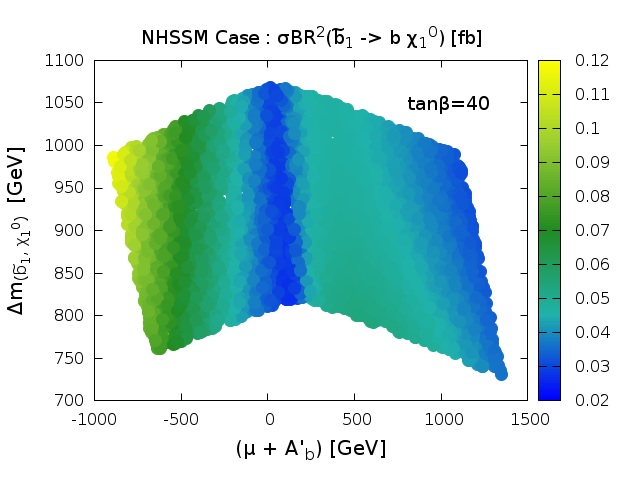}
	}
	\caption{
Parton-level yields
($\sigma_{\sbone\sbone} \times \mathrm{BR}[\sbone \to b \ntrlone]^2$) in the
final state $2b + \etmiss$ arising from pair-produced $\sbone$ at the 13 TeV run
of the LHC in the MSSM (left) and the NHSSM (right) for $\tanb=10$
(top panel) and 40 (bottom panel).
Magnitudes of the yield are indicated via the
color palettes
in the plane of the input parameter `$\mu$' ($\mu+\abprime$, for the plots on
right) and the resulting mass-split between $\sbone$ and the LSP,
$\Delta m_{(\sbone, \ntrlone)}$.
Other fixed parameters are: $\mu=200$ GeV, $\mone=500$ GeV,
$\mtwo=1.1$ TeV, $A_t=2$ TeV, $A_b=0$, $\msbleft$= $\msbright$=1.2 TeV and
$\muprime=0$.
\label{fig:sigma-br-b1}
}
\end{center}
\end{figure}

First, for simplicity, we consider $\muprime=0$. This is what we have already
adopted in sections \ref{subsec:coupling-sbot}
and \ref{subsec:masses}.
Furthermore, to stick to a
`natural' setup we restrict ourselves to relatively small values of `$\mu$'
($\lesssim 350$ GeV). The soft gaugino parameters  $\mone$ and $\mtwo$ are fixed
at 500 GeV and 1.1 TeV, respectively, thus rendering the two lighter
neutralinos and the lighter chargino higgsino-like\footnote{In practice, one
could have also considered a scenario with $\mone, \, \mtwo << \mu \, (\, \lesssim 500$
GeV) thus making the light electroweakinos gaugino-like but still the overall
scenario remaining somewhat ``natural''. However, in that case, one would not be
able to witness the key effect of an enhanced $y_b$ in the NHSSM scenario as the
decay of the bottom squarks would be mostly driven by appropriate gauge couplings.}.

In figure \ref{fig:sigma-br-b1} we present a comparative study of the
parton-level yields
($\sigma_{\sbone\sbone} \times \mathrm{BR}[\sbone \to b \ntrlone]^2$) in the
final state $2b + \etmiss$ arising from pair-produced $\sbone$ at the 13 TeV run
of the LHC in the MSSM (left) and the NHSSM (right) for $\tanb=10$ (top) and 40
(bottom). As mentioned earlier, the spectra, the branching ratios of various
new-physics excitations and values of other low-energy observables are computed
using {\tt SARAH}-generated {\tt SPheno} and the cross sections for the relevant
processes are calculated using {\tt MadGraph5\_aMC@NLO (v2.6.0)}
\cite{mad-launch, Alwall:2014hca} with its default setup. Magnitudes of the
yield are indicated via the color palettes in the plane of the input parameter
`$\mu$' ($\mu+\abprime$, for the plots on right) and the resulting mass-split
between $\sbone$ and the LSP, $\Delta m_{(\sbone, \ntrlone)}$. For the purpose, we
varied `$\mu$' and $\abprime$ over the following ranges:
$100 \, \mathrm{GeV} \leq \mu \leq 350 \, \mathrm{GeV}$ and
$|\abprime| \leq 1.2$ TeV. 
The former would ensure the scenario to be `natural'. The range of $\abprime$
ensures no
appearance of tachyonic sbottom states or CCB minima
of the scalar potential while causing appropriate scalar mixing that are
instrumental to
the present analysis. 

By comparing the left and the right plots in a given row ({\it i.e.}, for a
given $\tanb$) of figure \ref{fig:sigma-br-b1}, it is clear that the yield in
the NHSSM for a particular value of `$\mu$' ({\it i.e.},
$\mntrlone \simeq \mcharone$, for $\mu << \mone,\mtwo$) could vary widely as
opposed to a rather definite expectation for the same in the MSSM. This becomes
clear from the plots on the right in which the horizontal axes depict the
variation of the NHSSM construct $\mu+\abprime$ that appears in the
off-diagonal term of the symmetric sbottom mass-squared matrix and augments the
corresponding MSSM matrix for which only `$\mu$' appears. The vertical bands in
blue in the central region of the plots on the right with smaller values of
$\mu+\abprime$ are indicative of the smallest yields. These are commensurate
with a small mixing in the sbottom sector over this region. As discussed in
section \ref{subsec:coupling-sbot}, such a small mixing leaves $\sbone$
dominantly left-handed and hence with an enhanced BR[$\sbone \to t \charonem$].
Consequently, one finds a suppressed BR[$\sbone \to b \ntrlone$] over this
region. We have checked that the patterns shown in the plots on the right
closely follow the same for the above branching fraction. The latter, in the
first place, inherits its pattern from those of the interaction strengths as
illustrated in figure \ref{fig:couplings-abprime-higgsino}. Such a similarity is
seen for large negative values of $\mu+\abprime$
({\it i.e.}, for large negative $\abprime$) when we obtain the highest yield (in
yellow).

As for the MSSM case ({\it i.e.}, the plots on the left), it may further be
noted that the variations in
$\sigma_{\sbone\sbone} \times \mathrm{BR}[\sbone \to b \ntrlone]^2$, {\it i.e.},
the parton-level yields,
are primarily due to variations in BR[$\sbone \to b \ntrlone$] which, in turn,
are attributed to the changing chiral contents of $\sbone$ as `$\mu$' varies.
This is clear since the available phase space ($\Delta m_{(\sbone, \ntrlone)}$) for
the above decay remains to be large enough over the entire range of variation of
`$\mu$'. This also lends credence to the fact that the (counter-intuitive)
increase in the yield at lower values of $\Delta m_{(\sbone, \ntrlone)}$ ({\it i.e.},
decreasing phase space for the decay) is actually connected to the composition
of $\sbone$ in these regions, as discussed above. In summary, for a given set of
values of $\msbone$ and $\mntrlone$ where the latter is higgsino-like and
relatively light, the yields in the NHSSM could surpass their corresponding MSSM
expectations by large margins. However, the magnitudes of $\sigma \times
\mathrm{BR}$ reveals that these are somewhat sensitive to the LHC experiments
with 300 fb$^{-1}$ of integrated luminosity only when $\tanb$ is reasonably
large.

Next, we bring in $\muprime$ into the picture. As pointed out in section
\ref{sec:model}, this could give rise to a relatively heavier higgsino-like
neutralino ($\sim 1$ TeV) LSP without requiring `$\mu$' to be large
\cite{Ross:2016pml, Chattopadhyay:2016ivr}. This would then help avoid an
imminent tension with the notion of `naturalness'. In addition, such a
neutralino LSP is known to be a viable DM candidate that could explain
the observed relic density. In view of the LHC data routinely pushing up the
lower bounds on sparticle masses and that a bino-dominated neutralino LSP
overproduces the relic abundance, these together certainly offer an welcome
respite.
Furthermore, a non-vanishing $\muprime$ would not by itself affect the
electroweakino sector in the sense that the latter is blind to `$\mu$' or
$\muprime$ separately but inherits its properties only from the sum
$\mu+\muprime$. However, of `$\mu$' and $\muprime$, only the former enters the
sbottom sector, at least at the lowest order. Such a selectiveness is expected
to leave its trails in the masses and interactions involving these two sectors
and the way they are connected. These would then carry imprints of the NHSSM
scenario and could even shed light on the relative magnitude of `$\mu$' and
$\muprime$.
%
\begin{figure}[t]
\subfigure{
\includegraphics[width=7.1cm]{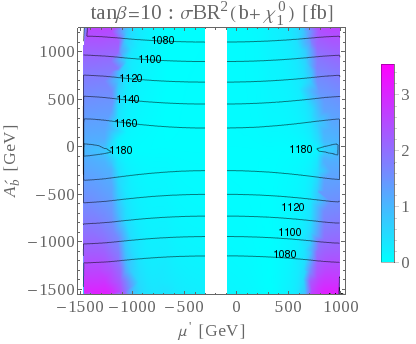}
}
\subfigure{
\includegraphics[width=7.5cm]{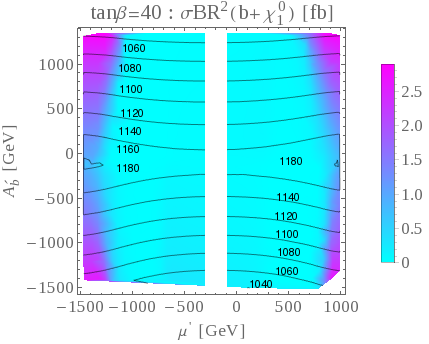}
}
\subfigure{
\includegraphics[width=7.4cm]{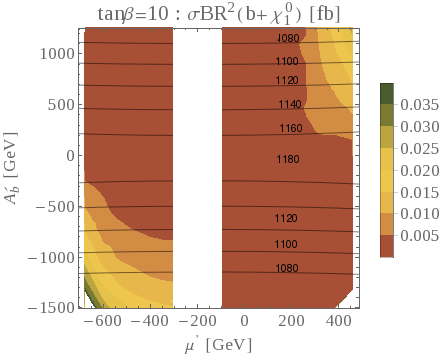}
}
\hskip 23pt
\subfigure{
\includegraphics[width=7.5cm]{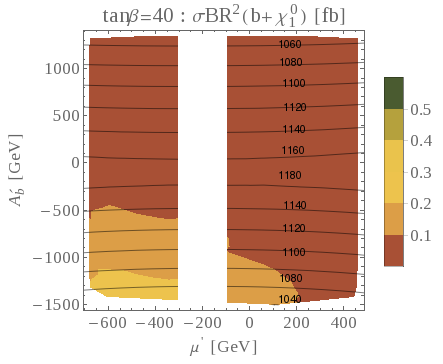}
}
\caption{Variation of $\sigma_{\sbone \sbone} \times \mathrm{BR}[\sbone \to b
\ntrlone]^2$ (yield) in the $2b + \etmiss$ final state as a function of
$\muprime$ and $\abprime$ for fixed MSSM configurations ($\mu=200$ GeV,
$\mone=500$ GeV, $\msbleft$= $\msbright$=1.2 TeV, $A_b=0$
with $\tanb=10$ (left) and 40 (right)) over an extended range of $\muprime$ (top
panel) and for a zoomed-up range with low $\muprime$ (bottom panel). The blank
vertical bands in the middle are roughly excluded by searches of the lighter
chargino at the LEP experiments, {\it i.e.}, $\mcharone \gtrsim 100$ GeV
\cite{lepsusywg}. See table \ref{tab:inputs} and the text for details.  
}
\label{fig:muprime-abprime}
\end{figure}

In figure \ref{fig:muprime-abprime} we present the possible extent of variation
of the yield ($\sigma \times \mathrm{BR}^2$) as a function of $\muprime$ and
$\abprime$ for $\msbleft$= $\msbright$=1.2 TeV.
This is done by fixing the MSSM parameters `$\mu$', $\mone$ and
$A_b$ in the following way: $\mu=200$ GeV, $\mone=500$ GeV and $A_b=0$. All
other input parameters are as indicated in table \ref{tab:inputs}. Note that the
chosen (small) value of `$\mu$' is expected to render the scenario `natural'.
All these, in turn, fix the MSSM yield. The plot on the left (right) corresponds
to $\tanb=10 \, (40)$.
By fixing $\mone$ and allowing for $\muprime$ to vary about it, we essentially
let the nature of the LSP neutralino change over from a dominantly higgsino-like
one to a gaugino-like one passing through an intermediate state of mixed nature
and thus, capture how the yield varies with such a change. Plots in the bottom
panel zoom in on the higgsino-like LSP region. The white bands in the middle of
the plots indicate regions which are excluded by the LEP chargino searches
\cite{lepsusywg}. Fixed $\msbone$ contours are also overlaid in each plot.

In both plots of the top panel of figure \ref{fig:muprime-abprime}, the regions
in cyan are of the smallest yields. The fixed MSSM yields for both $\tanb=10$
and 40 are obtained for $\muprime=\abprime=0$ and those are as small as 0.0003
fb and 0.039 fb, respectively. We have checked that the smallness of the yield
over the cyan region is mostly due to the same for the effective interaction
strength $C_L^2+C_R^2$. Even the presence of a slightly darker vertical shade in
the middle of the plot on the right indicating locally enhanced yields over the
range $-700 \, \mathrm{GeV} \lesssim \muprime \, \lesssim 400 \, \mathrm{GeV}$
and for $\abprime < 0$ can be traced back to similar enhancements in
$C_L^2+C_R^2$. Also, note that the magnitude of the yield varies in a
discontinuous fashion with increasing $|\muprime|$. This is due to an interplay
between the $y_b$-driven contribution which falls off more rapidly with growing
$|\muprime|$ (because of a decreasing higgsino content of the LSP) than what the
growing gaugino (bino) contribution could compensate for. Note that even though
the gaugino-dominance sets in at length for larger values of $|\muprime|$ and
takes control over the yield, its `extreme' values draw heavily (though
indirectly) from the ever-diminishing branching fraction
BR[$\sbone \to t \charonem$] as $\mcharone$ grows with $|\muprime|$, before
getting kinematically forbidden. In the bottom panel of figure
\ref{fig:muprime-abprime} we zoom up the low $|\muprime|$ region to illustrate
the altering nature of the yield and its extent across the region. It can be
gleaned from these plots that a 5 to 7 fold change in the yield is possible
over the indicated range.

In summary, an enhanced $y_b$, which is rather characteristic of the NHSSM
scenario for large negative $\abprime$ and large $\tanb$, could boost the yield
in the $2b+\etmiss$ final state beyond its MSSM expectation, for similar masses
of the lighter sbottom and the LSP in the two scenarios. The reverse is true
when $\abprime$ takes large positive values. In particular, the enhancement is
significant for a higgsino-like LSP and is somewhat more interesting when the
lighter sbottom is $\sbleft$-dominated given that there is a competition among
the decays $\sbone \to b \ntrlonetwo$ and $\sbone \to t \charonem$.
%
\subsubsection{Impact of including the heavier sbottom}
\label{subsubsec:lhc-sbtwo-pair}
%
A priori, it would not be fair to ignore the contribution from
$pp \to \sbtwo \sbtwo^*$. This is since, for the ranges of variations of various
parameters (like $\abprime$ and $\tanb$), $\msbone$ and $\msbtwo$ may not be too
different, in particular, when we consider $\msbleft$ and $\msbright$ to be
degenerate. It would thus be instructive to check what role could $\sbtwo$
possibly play in the phenomenology.

\begin{figure}[t]
\begin{center}
%
\raisebox{4mm}[0pt][0pt]{
\includegraphics[width=6.6cm,
height=6.0cm]{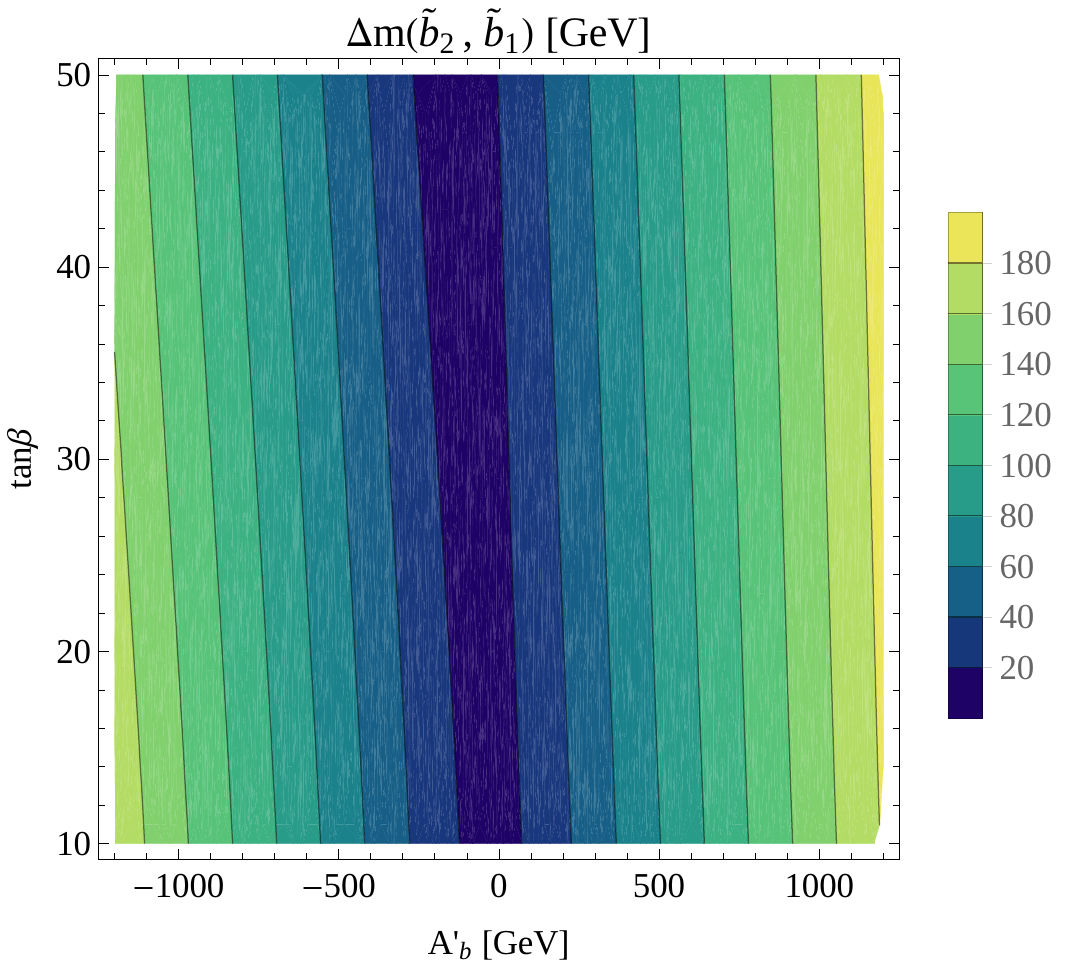}}
\hskip 20pt
\includegraphics[width=6.0cm, height=6.8cm]{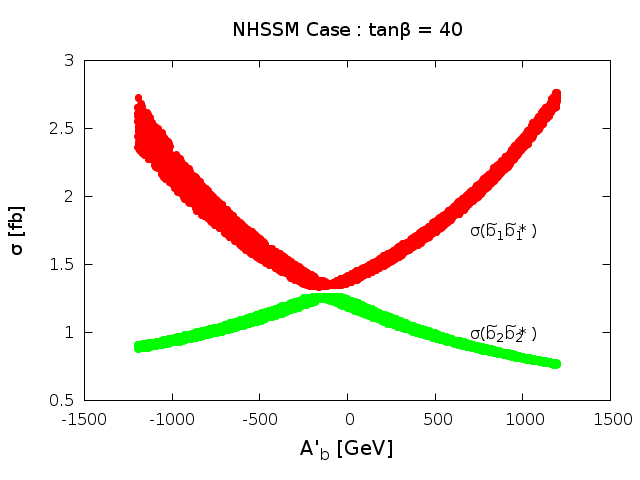}
\caption{Contours of constant $\Delta m_{(\sbtwo, \sbone)}$ in the
$\abprime$--$\tanb$ plane (left) and variations of
$\sigma(pp \to \sbone \sbone^*, \, \sbtwo \sbtwo^*)$ as functions of $\abprime$
for $\tanb=40$ and
$100 \; \mathrm{GeV} \leq \mu \leq 350$ GeV. For both plots
$\msbleft$= $\msbright$=1.2 TeV. Other fixed parameters are $A_b=0$, $\mu=200$ GeV,
$\muprime=0$, $A_t=2$ TeV, $\mone=500$ GeV and $\mtwo=1.1$ TeV.  
}
\label{fig:delta-msb1-msb2-sigma}
\end{center}
\end{figure}

To illustrate this, in the left plot of figure \ref{fig:delta-msb1-msb2-sigma}
we present the contours of constant mass-split ($\Delta m_{(\sbtwo, \sbone)}$)
between $\sbtwo$ and $\sbone$ in the $\abprime$--$\tanb$ plane. We note that for
the extreme value for $|\abprime|$ (=1.2 TeV) that we have allowed for in the
present analysis, the split between $\msbone$ and $\msbtwo$ cannot be more than
around 170 GeV. Furthermore, as can be seen in figure
\ref{fig:delta-msb1-msb2-sigma}, the mass-split is largely independent of
$\tanb$. Thus, for low values of $|\abprime|$, $\msbone$ and $\msbtwo$ could lie
very close by and hence $\sigma(pp \to \sbtwo \sbtwo^*)$ may turn out to be not
much smaller than
$\sigma(pp \to \sbone \sbone^*)$. These cross sections, as functions of
$\abprime$, are compared in the right plot of figure
\ref{fig:delta-msb1-msb2-sigma} for a fixed value of $\tanb=40$. Bands arise due
to variation of `$\mu$' over the range $100 \, \mathrm{GeV} \leq \mu \leq 350$
GeV. We have checked that the pattern of variation or the values of cross
sections do not alter much if a different value of
$\tanb$ is chosen. This is not unexpected since the production cross sections
are mainly governed by the masses of the sbottoms. What then could matter for
the yield in the final state of our interest, {\it i.e.}, $2b+\etmiss$,
is BR[$\sbtwo \to b \ntrlone$].
%
%
%
\begin{figure}[t]
\begin{center}
\includegraphics[width=7.5cm]{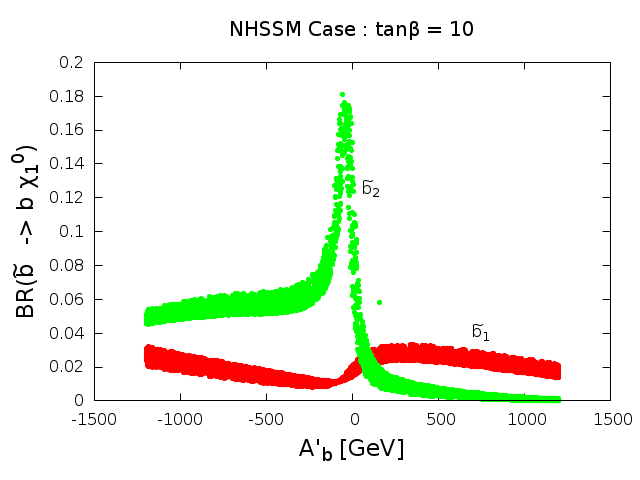}
\hskip 20pt
\includegraphics[width=7.5cm]{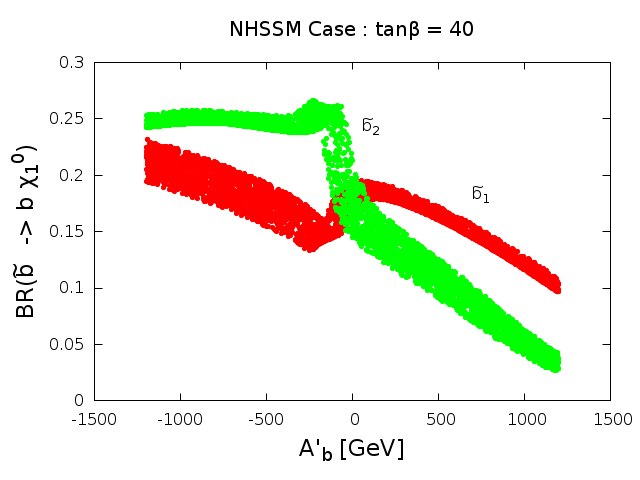}
\caption{Branching fractions of $\sbone$ ($\sbtwo$) to $b \ntrlone$ shown in
red (green) as functions of $\abprime$ (GeV). The left (right) panel stands for
$\tanb=10 \; (40)$. Other fixed parameters are:
$\mone=500$ GeV, $\mtwo=1.1$ TeV, $A_t=2$ TeV, $A_b=0$,
$\msbleft$= $\msbright$=1.2 TeV and $\muprime=0$ with $100 \leq \mu \leq 350$ GeV. 
}
\label{fig:b1-b2-BR-LSP}
\end{center}
\end{figure}

In figure \ref{fig:b1-b2-BR-LSP} we present the variations of branching
fractions BR[$\sbone \to b \ntrlone$] and BR[$\sbtwo \to b \ntrlone$] as
functions of $\abprime$ for $\tanb=10$ and 40. Unlike the variations of their
pair-production cross sections, these show some significant quantitative
difference between the cases with $\tanb=10$ and 40. As can be seen,
BR[$\sbtwo \to b \ntrlone$] is always larger than
BR[$\sbone \to b \ntrlone$] for $\abprime < 0$. The largest difference is seen
around vanishing $\abprime$ where BR[$\sbtwo \to b \ntrlone$] peaks while
BR[$\sbtwo \to b \ntrlone$] touches the minimum. The phenomenon could be
understood in terms of the sharply increasing dominance of $\sbright$ in
$\sbtwo$ as $|\abprime| \leadsto 0$. This quickly suppresses
BR[$\sbtwo \to t \charonem$] in favor of BR[$\sbtwo \to b \ntrlone$].
The situation is just the opposite in the case of
BR[$\sbone \to b \ntrlone$] thus leading to the contrast seen for the two
variations. Such an enhanced magnitude of BR[$\sbtwo \to b \ntrlone$] for
$\abprime <0$ could very well compensate for a relatively smaller value of
$\sigma(pp \to \sbtwo \sbtwo^*)$. Thus, the yield in the $2b+ \etmiss$ final
state from $\sbone \sbone^*$ and $\sbtwo \sbtwo^*$ could be comparable and
hence the latter should not be ignored.  
%
\begin{figure}[t]
\begin{center}
\includegraphics[width=7.5cm]{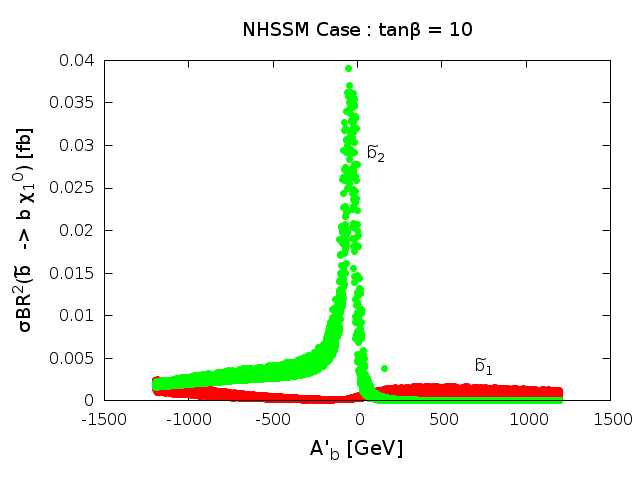}
\hskip 20pt
\includegraphics[width=7.5cm]{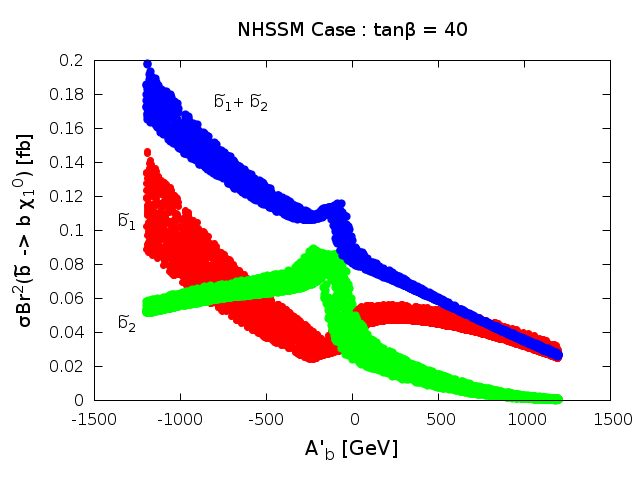}
\caption{Variation of parton-level yields in the $2b+\ntrlone$ final state
arising from $\sbone$ (in red) and $\sbtwo$ (in green) pair-production at the
13 TeV LHC as functions of $\abprime$. The left (right) panel stands for
$\tanb=10 \, (40)$. Other fixed/varying parameters are as in figure \ref{fig:b1-b2-BR-LSP}.
For the plot on the right, contributions from $\sbone$ and $\sbtwo$ are added up
(in blue).
}
\label{fig:sigma-br-abp}
\end{center}
\end{figure}
%
\begin{figure}[h]
\begin{center}
\includegraphics[width=7.5cm]{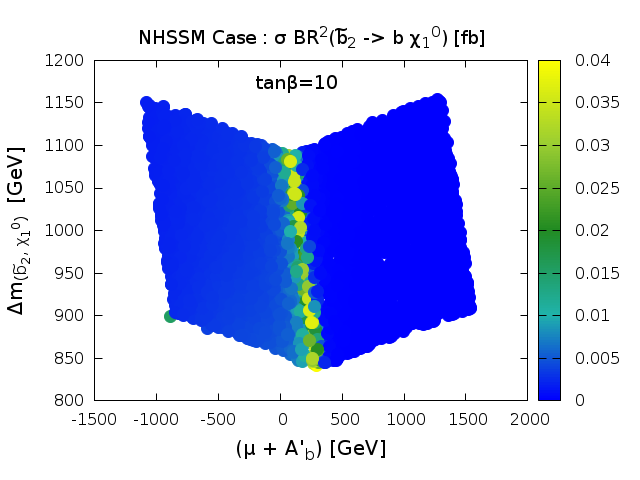}
\hskip 20pt
\includegraphics[width=7.5cm]{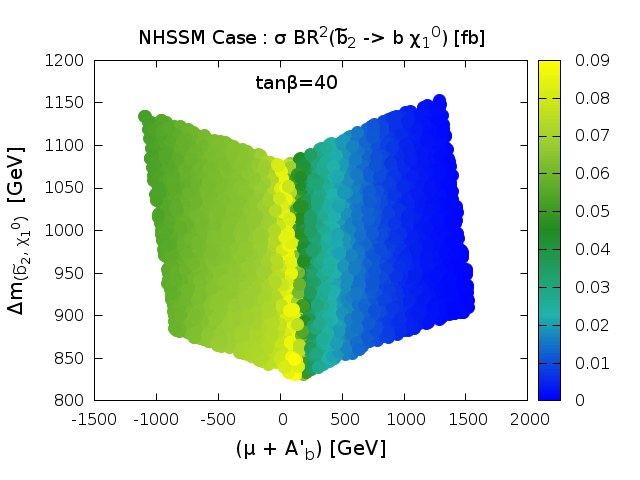}
\caption{Parton-level yields ($\sigma_{\sbtwo \sbtwo^*} \times
\mathrm{BR}[\sbtwo \to b \ntrlone]^2$) in the final state $2b + \etmiss$ arising
from pair-produced $\sbtwo$ at the 13 TeV run of the LHC in the NHSSM for
$\tanb=10$ (left panel) and 40 (right panel) in the plane of $\mu+\abprime$
and the resulting mass-split between $\sbtwo$ and the LSP, $\Delta m_{(\sbtwo,
\ntrlone)}$. Magnitudes of the yield are indicated via color palettes.
Other fixed/varying parameters are as in figure \ref{fig:b1-b2-BR-LSP}.
}
\label{fig:sigma-br-b2}
\end{center}
\end{figure}

In figure \ref{fig:sigma-br-abp} we show the variations of the parton-level
yields ({\it i.e.}, the cross section times the square of
BR[$\tilde b_{1/2} \to b \ntrlone$]) in the $2b + \etmiss$ final state
arising from $\sbone$- (in red) and $\sbtwo$ (in green) pair-productions for
$\tanb=10 \, (40)$ in the left (right) plot. For the plot on the
right, we also added up these two contributions (in blue) as in this case the
individual contributions are relatively large over the range we explored.
As one can see, for small values of $|\abprime|$, yield
from $\sbtwo$ pair-production dominates and this simply inherits its trend
from the plot on the right of figure \ref{fig:delta-msb1-msb2-sigma} and from figure
\ref{fig:b1-b2-BR-LSP}. However, with
small $|\abprime|$ the scenario tends to become MSSM-like over this region.
Interestingly, for relatively large negative $\abprime$ and for large $\tanb$
the combined contribution from $\sbone$ and $\sbtwo$ pair-production could
exceed the MSSM expectation significantly. For a given set of
values of $\msbone$, $\msbtwo$ and $\mntrlone$, this then could signal in favour of
a scenario like the NHSSM with a few tens of events in 300 \fbinv of data at
the LHC. On the other hand, for large positive $\abprime$, one expects a
significant dearth of events in the $2b+\etmiss$ final state when compared to the
MSSM expectation. However, the yield
is found to be not sensitive to 300 \fbinv of data and would need the
high luminosity run of the LHC for this to show up.
In figure \ref{fig:sigma-br-b2} we illustrate the variation of
$\sigma_{\sbtwo\sbtwo^*} \times \mathrm{BR}[\sbtwo \to b \ntrlone]^2$ in the
$(\mu+\abprime)$--$\Delta m_{(\sbtwo, \ntrlone)}$ plane for $\tanb= 10$ (left) and
40 (right). These correspond to the similar ones for $\sbone$ presented in the
right plots in the top and bottom panels of figure \ref{fig:sigma-br-b1}.

\begin{figure}[t]
\begin{center}
\includegraphics[width=6.5cm,height=4.5cm]{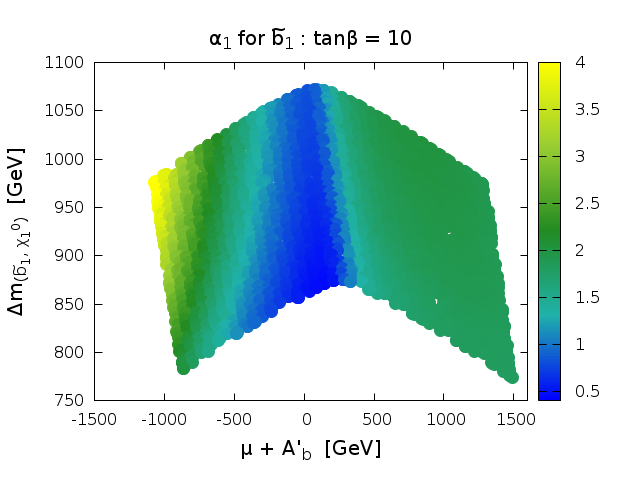}
\hskip 20pt
\includegraphics[width=6.5cm,height=4.5cm]{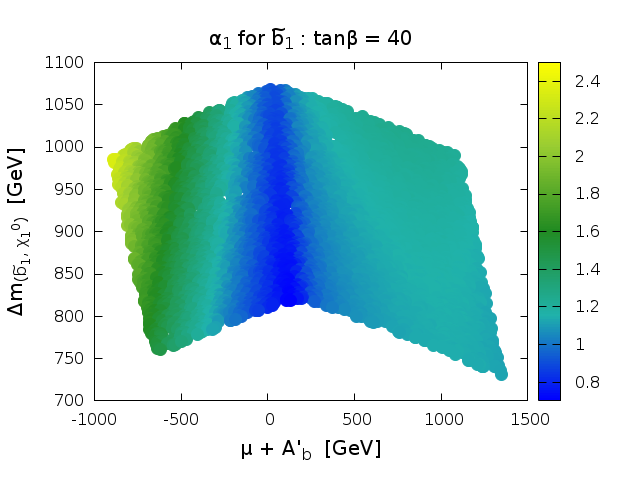}
\vskip 10pt
\includegraphics[width=6.5cm,height=4.5cm]{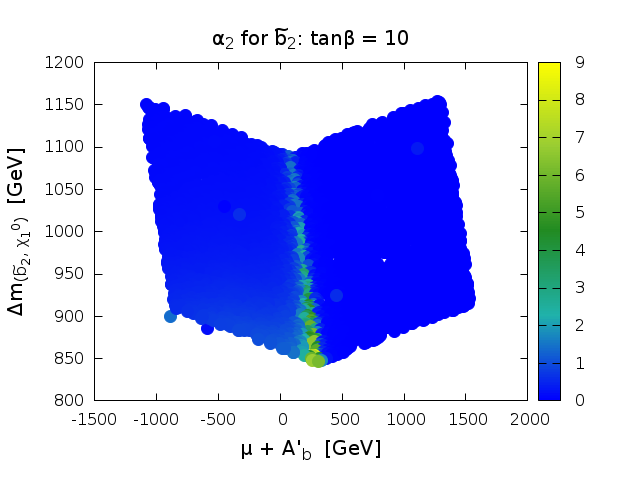}
\hskip 20pt
\includegraphics[width=6.5cm,height=4.5cm]{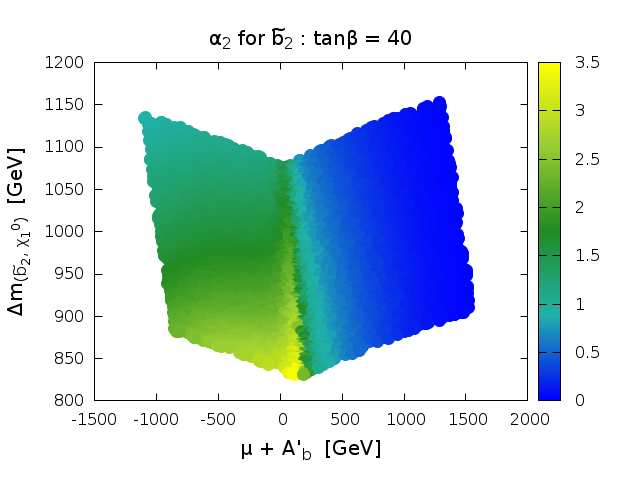} 
\vskip 10pt
\includegraphics[width=6.5cm,height=4.5cm]{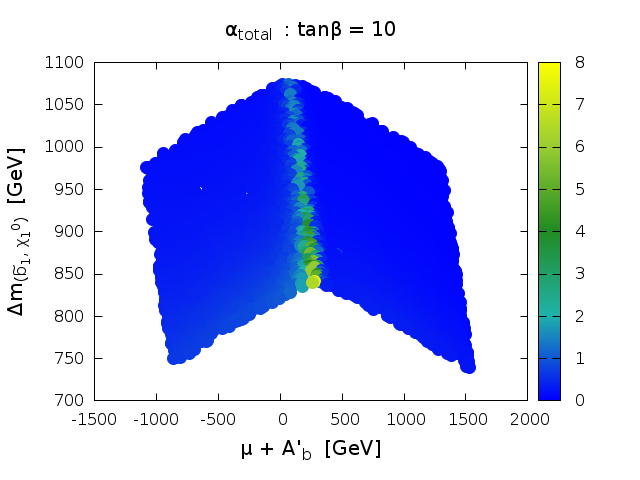}
\hskip 20pt
\includegraphics[width=6.5cm,height=4.5cm]{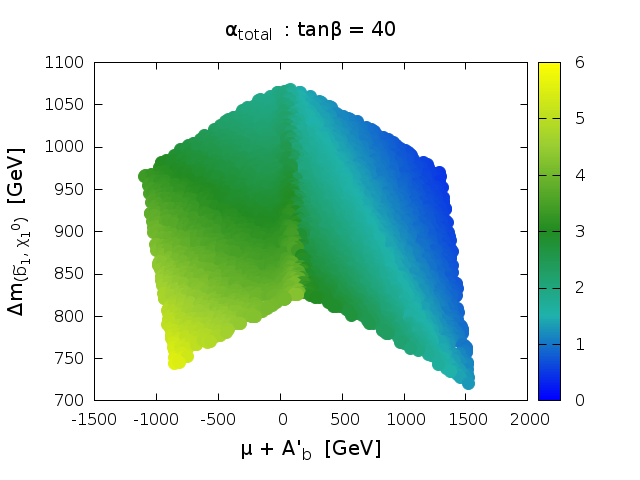} 
\caption{Relative rates ($\alpha$) between the NHSSM and the MSSM scenarios in
the final state $2b+\etmiss$ in the plane of $\mu+\abprime$ and
$\Delta m_{(\tilde{b}_i, \ntrlone)}$ obtained from $\sbone$ pair-production (top row), $\sbtwo$
pair-production (middle row) and sum of $\sbone$ and $\sbtwo$ pair-production
(bottom row) for $\tanb=10$ (left panel) and 40 (right panel).
`$\mu$' is varied over the range $100 \, \mathrm{GeV} \leq \mu \leq 350$ GeV
while the other fixed parameters are: $\mone=500$ GeV, $\mtwo=1.1$ TeV,
$A_t=2$ TeV, $A_b=0$, $\msbleft$= $\msbright=$1.2 TeV and $\muprime=0$.
}
\label{fig:alpha}
\end{center}
\end{figure}

We now summarize our findings
by undertaking a
simple-minded comparison of the $2b+\etmiss$ rates obtained in the MSSM and in
the NHSSM, as $\abprime$ varies for the same values of $\mu$ ({\it i.e.}, similar
higgsino-like LSP masses). We define the relative rates as
%
\begin{equation}
\alpha_i (\abprime) = \frac{\big[(\sigma_{\tilde{b}_i \tilde{b}_i} \times
\mathrm{BR}[\tilde{b}_i \to b \tilde{\chi}^0_1]^2 )
\big]^{\mathrm{NHSSM}} }
{\big[(\sigma_{\tilde{b}_i \tilde{b}_i} \times
\mathrm{BR}[\tilde{b}_i \to b \tilde{\chi}^0_1]^2 )
\big]^{\mathrm{MSSM}}}
\quad \mathrm{and} \quad
\alpha_{\mathrm{total}}(\abprime) = \frac{\sum\limits_{i=1,2} \big[(\sigma_{\tilde{b}_i \tilde{b}_i} \times
\mathrm{BR}[\tilde{b}_i \to b \tilde{\chi}^0_1]^2 )
\big]^{\mathrm{NHSSM}}}
{\sum\limits_{i=1,2} \big[(\sigma_{\tilde{b}_i \tilde{b}_i} \times
\mathrm{BR}[\tilde{b}_i \to b \tilde{\chi}^0_1]^2 )
\big]^{\mathrm{MSSM}}}
\label{eq:alpha}
\end{equation}
%
\vskip 10pt
\noindent
with $\msbleft$=$\msbright$ = 1.2 TeV and its variation is shown in
figure \ref{fig:alpha} in the plane of $\mu+\abprime$ and
$\Delta m_{(\sbone, \ntrlone)}$. We obtain the
rates in the MSSM case by varying `$\mu$' over the range
$100 \, \mathrm{GeV} \leq \mu \, \leq 350 \, \mathrm{GeV}$ such that a
relatively light higgsino-like neutralino LSP is ensured. For the NHSSM case, we
choose $\muprime=0$ but vary $\abprime$ such that $|\abprime| \leq 1.2$ TeV to 
get the yields. To compute `$\alpha$', the ratios are taken of those rates in
the two scenarios for which values of `$\mu$' are the same. The plots
reveal that up to a eight-fold (six-fold) increased rates could be possible for
$\tanb=10 \, (40)$ over the expected MSSM rates in the final state under
consideration. These further indicate that NHSSM could also result in a much
lower yield when compared to the MSSM. Note, however, that the absolute rates
are typically lower for a smaller $\tanb$. Understandably, the largest deviation
is expected for relatively large, negative values of $\abprime$ for which $y_b$
is much enhanced. It should be noted that the variations of `$\alpha$' closely
mimics that of $\sigma \times \mathrm{BR}^2$ in figure \ref{fig:sigma-br-b1} and
\ref{fig:sigma-br-b2}. Hence the former variations find very similar explanations
in terms of variation of the effective interaction strengths as well.

\begin{figure}[t]
 \begin{center}
  \subfigure{
  \includegraphics[width=7.8cm]{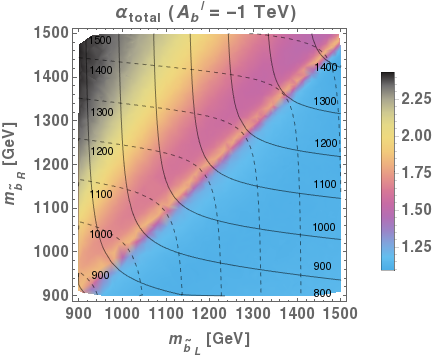}
  }
  \subfigure{
  \includegraphics[width=7.8cm]{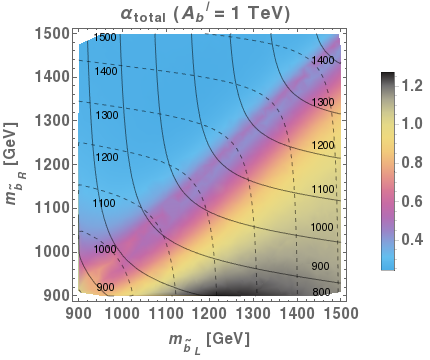}
  }
\caption{Variations of $\alpha_{\mathrm{total}}$ in the NHSSM
and in the MSSM in the $\msbleft - \msbright$ plane for
$\abprime=-1$ TeV (left) and $\abprime=1$ TeV (right) and for fixed values of
$\tanb$ (=40) and $\mu$ (=200 GeV). Other fixed parameters are as in figure
\ref{fig:alpha}. Contours of constant $\msbone$ ($\msbtwo$) are overlaid with
solid (dashed) lines along the right (left) edges of the plots.
}
\label{fig:ratioR}
\end{center}
\end{figure}

Finally, it will be of practical importance to understand how the rates would
compare when the masses of the sbottoms vary. To this end, in figure
\ref{fig:ratioR} we illustrate the
variations of a similar ratio of the expected (total) yields as defined in
equation \ref{eq:alpha} (but for two representative choices of $\abprime$ and
$\tanb=40$) in the plane
of $\msbleft$ and $\msbright$. Resulting contours of fixed $\msbone$ and
$\msbtwo$ are overlaid.
One finds that for the plot on left with $\abprime=-1$ TeV, the maximal
deviation in the yields in the $2b$-jets+$\etmiss$ final state with respect to
the MSSM yield occurs for $\sbone$-dominated by $\sbleft$ (black region).
The reverse is true for the plot on the right with $\abprime=1$ TeV. Both of
these situations could be understood in terms of the way $y_b$ varies with
$\abprime$ and the added role $\abprime$ plays in the mixing in the sbottom
sector in the NHSSM, over and above the already existing MSSM effects. 

To elaborate a bit more, for $\abprime=-1$ TeV as is used in the left plot,
we now know that $y_b$ is significantly larger than its MSSM value. As the decay
$\tilde{b} \to b \ntrlone$ is enhanced for larger $y_b$, we always find the relative
yield to be larger than 1. It turns out that the maximal deviation (black region)
occurs when the MSSM rate (appearing in the denominator of the ratio) in the
said channel becomes minimum which is the case when $\sbone \approx \sbleft$.
This is so since such a limit is more abruptly attained in the MSSM than in
the NHSSM thanks to a non-vanishing $\abprime$ for which $\sbone$ could still
have a larger $\sbright$ admixture in the NHSSM case which facilitates
the decay $\sbone \to b \ntrlone$. In moving to the plot on the right, the MSSM
contribution in the denominator does not change. However, with $\abprime=1$ TeV,
$y_b$ becomes much smaller than its MSSM value. Thus, in the region for which $\sbone$
becomes dominated by $\sbleft$ (left of the diagonal), $\sbone \to t \charonem$
prevails. This is the reason the ratio of the yields are now smaller over this
region. The maximal ratio (black patch) is now obtained for $\sbone \approx
\sbright$ and this appears entirely due to enhanced cross section in the NHSSM
case where $\sbone$ gets lighter than that in the case of the MSSM for the same
input values of $\msbleft$ and $\msbright$. Note that this effect is also there
over the black patch in the left plot which gets further fortified by the
effects discussed above. Clearly, these can turn the ratio bigger as can be seen
for $\abprime < 0$.   
%
%
\begin{figure}[t]
 \begin{center}
  \subfigure{
  \includegraphics[width=7.8cm]{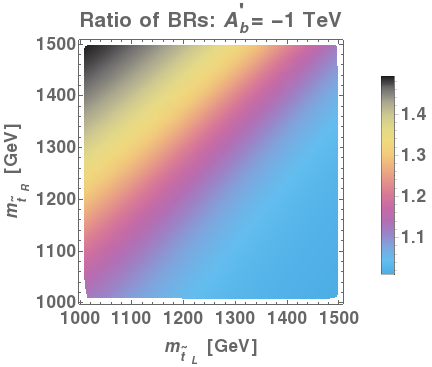}
  }
  \subfigure{
  \includegraphics[width=7.8cm]{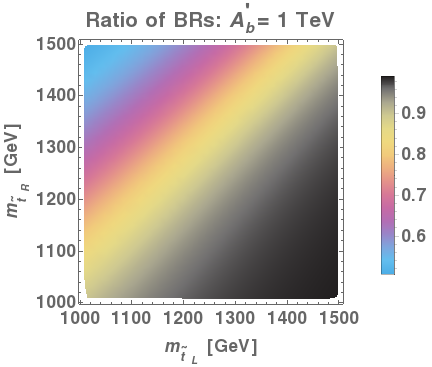}
  }
\caption{Variations of the ratio of branching fractions for the decay
$\stone \to b \charonep$ in the NHSSM and the MSSM in the $\stleft$-$\stright$
plane for $\abprime=-1$ TeV (left) and $\abprime=1$ TeV (right) and for fixed
values of $\tanb$ (=40) and $\mu$ (=200 GeV). Other fixed parameters are as in
figure \ref{fig:ratioR}.
}
\label{fig:stopBR}
\end{center}
\end{figure}
%
\subsection{Implication for stop searches}
\label{subsec:lhc-stop}
%
In this subsection we briefly discuss possible implications of the NHSSM
scenario for the stop searches at the LHC. It has been pointed out in the
Introduction that the impact of soft NH terms on the masses and mixings of
the stops cannot be large since this is going to be $\tan\beta$-suppressed.
However, as indicated there, it remains to be seen if an altered $y_b$ could play
any role in the decays of stops. Indeed, an expression analogous to
equation \ref{eqn:sbtc1} for the  
$\stopi$-$b$-$\charonep$ vertex would dictate that
a left-like stop would couple to a higgsino-like chargino and a bottom quark
with the strength $y_b$. Hence its decay rate to $b \charonep$ could get
modified when compared to its MSSM expectation depending upon the altered value
of $y_b$ as functions of $\abprime$ and $\tanb$.

In figure \ref{fig:stopBR} we illustrate the variations of the relative
branching rate of $\stone$ to $b\charonep$ in the NHSSM (with respect to the
MSSM case) in the plane of $\stleft$ and $\stright$ for two values of $\abprime$
(-1 TeV (left) and 1 TeV (right)) and for $\tanb=40$ for which the changes in
the values of $y_b$ are moderately large. It is clear from these plots that 
when $\stone$ is $\stleft$-dominated ({\it i.e.}, when $\mstleft << \mstright$),
the effect of altered $y_b$ is maximal. For $\abprime < 0$ (left plot) leading
to $y_b$ which is larger than its MSSM expectation, the effect is manifested in
an enhanced (the black patch) branching fraction in the top left region. The
reverse is true for the right plot where the said region now suffers the most
due to a diminished $y_b$ (the blue patch) as is expected for $\abprime > 0$. 
%
\section{Conclusions}
\label{sec:conclusions}
%
Deciphering imprints of the NHSSM scenario at colliders would not be a simple
proposition. Given the way the two important classes of nonholomorphic soft
terms in the form of $\muprime$ and $A_f'$ appear in the NHSSM Lagrangian,
perhaps the only plausible way to extract information about them is to undertake
a thorough, precise study of the interactions of the sfermions with the
electroweakinos.

In the present work, we mostly adopt a scenario in which the SUSY
conserving parameter `$\mu$' has a relatively small value ($\leq 350$ GeV) which
help keep the scenario `natural'. Furthermore, we prefer the masses of the
low-lying electroweakinos to be governed by `$\mu$' and hence turning out to be
higgsino-like. This then could exploit the dominant Yukawa couplings thus bringing
the sfermions from the third generation into the folds, which could potentially
be the lightest of the sfermions and hence could be within the reach of the LHC.

We find the sbottom sector to be especially sensitive to the NHSSM soft terms
thanks to a very prominent dependence of the bottom Yukawa coupling $y_b$ on
the trilinear soft NH parameter $\abprime$ due to radiative
effects, reinforced by possible large values
of $\tanb$. This could have a moderate to large (and hence phenomenologically
important) effect on the masses and the mixings of sbottoms. For example, while
in the MSSM, the mass-split between the two sbottom states could only be around
1-2\% of the degenerate chiral soft masses in the most favorable scenario, one
could achieve a 10-15\% split in the NHSSM. In addition, the sbottoms, when
produced in pairs at the LHC, might lead to a $2b+\etmiss$ final state via
one-step decays of the sbottoms. Such a final state is being intensively looked
for at the LHC experiments. This is in contrast to pair-produced light stops
leading to top quarks or charginos in their decays both of which, in turn,
undergo cascades giving rise to myriad possibilities in the final state
involving intricate model-dependencies and in cases, having involved SM backgrounds.

In this work, we study in much detail the couplings of the sbottoms with the
electroweakinos which exploit to the fullest the simultaneous dependence of
$y_b$ on $\abprime$ and $\tanb$ along with benefiting from specific chiral
imprints of the sbottom states. For large, negative $\abprime$ and large
$\tanb$, we demonstrate that the NHSSM scenario could lead to a much healthier
event rate in the $2b+\etmiss$ final state when compared to the MSSM for similar
masses for the sbottom(s) and the LSP neutralino. The reverse is also possible
for large positive $\abprime$ thus ending up with a lower event count than what
is expected in the MSSM. We find that under favorable circumstances, with
sbottom masses not exceeding $\sim 1.5$ TeV, 300 \fbinv of LHC data could be
sensitive to such excesses. On the other hand, we note that a possible depletion
in the event count could only be convincingly established by the high luminosity
run of the LHC. A suitably designed multi-channel study could prove to be more
efficient in search for a powerful discriminator in the present exercise. 

Last but not the least, we have demonstrated how the dependence of $y_b$ on
$\abprime$ could have an important bearing on the phenomenology of the stops at
the LHC due to its altered branching fractions triggered by modified $y_b$.  
%
\acknowledgments
AD acknowledges the hospitality of the School of Physical Sciences, Indian
Association for the Cultivation of Science, Kolkata on several occasions during
the course of the present work. SM acknowledges the hospitality of Harish-Chandra
Research Institute, Allahabad during a collaborative visit during the same
period. AKS acknowledges the support received from Department of Science and
Technology, Government of India with fellowship (SERB NPDF) reference number
PDF/2017/002935.
%

%
\end{document}